\documentclass[aps,prb,twocolumn,floatfix,showkeys,superscriptaddress,amssymb,preprintnumbers]{revtex4-2}

\usepackage{graphicx}
\usepackage{bm}                   
\usepackage{amsmath}
\usepackage{csquotes}        
\usepackage[linktocpage, colorlinks, linkcolor = blue, citecolor = blue,urlcolor=blue]{hyperref}
\usepackage{dcolumn}        
\usepackage[dvipsnames, table]{xcolor}				
\usepackage{nicefrac}
\usepackage{multirow}
\usepackage{color, xcolor}

\usepackage{colortbl} 
\usepackage{booktabs} 
\usepackage{bigstrut} 
\usepackage{textcomp}
\newcommand\T{\rule{0pt}{2.6ex}}       
\newcommand\B{\rule[-1.2ex]{0pt}{0pt}}
 
\usepackage{soul} 
\usepackage{color, xcolor} 
\usepackage[colorinlistoftodos,textwidth=2cm]{todonotes}

\soulregister{\cite}7 
\soulregister{\citep}7 
\soulregister{\citet}7 
\soulregister{\ref}7 
\soulregister{\pageref}7 

\usepackage[absolute]{textpos}
\textblockorigin{10mm}{10mm} 
\begin{document}

\title{Lattice distortions and non-sluggish diffusion in BCC refractory high entropy alloys }

\author{Jingfeng Zhang}
\affiliation{School of Material Science and Engineering, Dongguan University of Technology, Dongguan 523808, China}
\affiliation{Institute of Materials Physics, University of M\"unster, Wilhelm-Klemm-Str. 10, M\"unster 48149, Germany}
\author{Xiang Xu}
\affiliation{Institute for Materials Science, University of Stuttgart, Pfaffenwaldring 55, Stuttgart 70569, Germany}
\author{Fritz K$\rm{\ddot{o}}$rmann}
\affiliation{Institute for Materials Science, University of Stuttgart, Pfaffenwaldring 55, Stuttgart 70569, Germany} 
\affiliation{Interdisciplinary Centre for Advanced Materials Simulation, Ruhr-Universit{\"a}t Bochum, 44801 Bochum, Germany}
\affiliation{Department for Computational Materials Design, Max-Planck-Institut for Sustainable Materials,  Max-Planck-Str.1, 40237 D$\ddot{u}$sseldorf, Germany}
\author{Wen Yin}
\affiliation{Institute of High Energy Physics, Chinese Academy of Science, Dongguan 523000, China}
\affiliation{Spallation Neutron Source Science Center, Dongguan 523803, China}
\author{Xi Zhang}
\affiliation{Institute for Materials Science, University of Stuttgart, Pfaffenwaldring 55, Stuttgart 70569, Germany}
\author{Christian Gadelmeier}
\affiliation{Metals and Alloys, University of Bayreuth, Prof.-R\"udiger-Bormann-Str. 1, Bayreuth 95447, Germany}
\author{Uwe Glatzel}
\affiliation{Metals and Alloys, University of Bayreuth, Prof.-R\"udiger-Bormann-Str. 1, Bayreuth 95447, Germany}
\author{Blazej Grabowski}
\affiliation{Institute for Materials Science, University of Stuttgart, Pfaffenwaldring 55, Stuttgart 70569, Germany}
\author{Runxia Li*}
\affiliation{School of Material Science and Engineering, Dongguan University of Technology, Dongguan 523808, China}
\affiliation{Dongguan Institute of Science and Technology Innovation, Dongguan University of Technology, Dongguan 523808, China}
\author{Gang Liu}
\affiliation{State Key Laboratory for Mechanical Behavior of Materials, Xi’an Jiaotong University, Xi’an 710049, China}
\author{Biao Wang}
\affiliation{School of Material Science and Engineering, Dongguan University of Technology, Dongguan 523808, China}
\author{Gerhard Wilde}
\affiliation{Institute of Materials Physics, University of M\"unster, Wilhelm-Klemm-Str. 10, M\"unster 48149, Germany}
\author{Sergiy V. Divinski*}
\affiliation{Institute of Materials Physics, University of M\"unster, Wilhelm-Klemm-Str. 10, M\"unster 48149, Germany}
\date{\today}

\begin{abstract}

Refractory high-entropy alloys (RHEAs) have emerged as promising candidates for extreme high-temperature applications, for example, in next-generation turbines and nuclear reactors. In such applications, atomic diffusion critically governs essential properties including creep resistance and microstructural stability. The present study systematically investigates impurity diffusion of Co, Mn, and Zn in single phase (BCC solid solution) HfTiZrNbTa and HfTiZrNbV RHEAs applying the radiotracer technique. A neutron total scattering technique is used to evaluate the pair distribution functions and element-specific lattice distortions in these alloys. \textit{Ab initio}-based calculations give access to lattice distortions and solubilities of the impurities under investigation, including the impact of short-range order. The diffusion results are discussed in relation to calculated substitutional and interstitial solution energies, local lattice distortions, and short-range order effects. Co diffusion is found to be dominated by the interstitial mechanism, exhibiting fast diffusion. These findings reveal important structure-property relationships between local atomic environments and diffusion kinetics in BCC RHEAs, providing critical insights for designing alloys with enhanced high-temperature performance through targeted control of impurity diffusion processes.

\end{abstract}


\maketitle

\begin{textblock}{3}(0,0)
	\framebox[1.01\width]{The paper is published in: Acta Materialia 297 (2025) 121283.  \url{https://doi.org/10.1016/j.actamat.2025.121283}~~} \par
\end{textblock}

\section{Introduction}

Refractory high-entropy alloys (RHEAs) have garnered substantial research interest since their initial development in 2010 \cite{senkov2010refractory}. The increased attention stems not only from the common four ``core effects'' associated generally with high-entropy alloys (HEAs), i.e. the high-entropy effect, ``sluggish'' diffusion, the cocktail effect, and severe lattice distortions \cite{Yeh2004}, but also from their exceptional strength at elevated temperatures, particularly within the range from 1200 to $1600^\circ$C \cite{senkov2010refractory,senkov2018development}. Among RHEAs, those composed of elements from groups IV (Ti, Zr, and Hf) and V (V, Nb, and Ta) stand out for their remarkable combination of high-temperature strength and excellent tensile plasticity. Notably, this impressive tensile plasticity persists even at ambient temperatures, a phenomenon predominantly attributed to severe lattice distortions inherently present in a solid solution of elements with significantly different atomic radii \cite{cook2024kink,gou2023additive,lei2018enhanced,senkov2018development,lee2018lattice,lee2020lattice,thirathipviwat2022role}.

However, the limited environmental resistance remains one of the most critical limitations for high-temperature structural applications of RHEAs \cite{senkov2018development}. RHEAs \cite{sheikh2018accelerated,muller2019oxidation,butler2019native,gorr2017high,wang2020effect} share a similar behavior with many refractory alloys \cite{distefano2000oxidation}, being susceptible to the formation of non-protective oxides. Internal oxidation or even a catastrophic failure, the so-called “pest” phenomenon, are observed in some refractory alloys with Al or Si additions \cite{sheikh2018accelerated,westbrook1964pest,sheikh2018aluminizing}. This vulnerability is primarily attributed to the rapid diffusion of oxygen through refractory metal oxides, facilitating the formation of new oxide layers beneath the existing scales. Such a susceptibility of RHEAs to rapid oxide growth seems to contradict the ``sluggish'' diffusion concept, one of the four core effects of HEAs, even though the diffusion of oxygen through the already formed oxide is most probably the rate-dominating process. In fact, direct experimental studies have demonstrated that self-diffusion of Zr in the TiZrHfNbV and TiZrHfNbTa RHEAs is not sluggish but is rather enhanced, if the geometric mean of diffusivities in the corresponding unaries (pure elements) is used as the reference \cite{zhang2022zr}. This definition of the diffusion sluggishness in multi-principal element alloys was further supported by large-scale molecular dynamics simulations \cite{Starikov2024}.

Since the introduction of the ``sluggish'' diffusion concept in high-entropy alloys in 2013 \cite{tsai2013sluggish}, extensive diffusion research has been performed, mainly on the most widely studied CoCrFeMnNi (i.e., ``Cantor'') HEA system \cite{gaertner2019concentration,zhang2021tracer,lukianova2020impact,gaertner2020tracer,kottke2020experimental,kottke2019tracer,vaidya2018bulk,vaidya2016ni,glienke2020grain,vaidya2017radioactive,choi2024grain,sen2024grain}. The findings reveal that the diffusion behavior in the Cantor alloy and related alloy systems is particularly complex and cannot be understood simply in terms of sluggish diffusion. Moreover, several diffusion studies in HCP HEAs have reported instances of rapid (or even ``anti-sluggish'') diffusion \cite{vaidya2020phenomenon,sen2023anti,sen2023does,sen2024sc}. While the diffusion characteristics vary strongly across different high-entropy alloys, recent studies suggest that the degree of lattice distortions significantly influences the diffusion behavior \cite{zhang2022zr,lukianova2022self,sen2023anti,daw2021sluggish}.  It should be noted here that the lattice distortions arise not only from the varying atomic sizes and electronic environments of the constituent elements of HEAs but also from the presence of small interstitial impurity atoms. For instance, distortions induced by small amounts of interstitial carbon significantly influence substitutional self-diffusion in the CoCrFeMnNi HEA \cite{lukianova2020impact,lukianova2022self}.

A conventional approach to quantifying lattice distortions involves calculating atomic size mismatch based on atomic radii \cite{zhang2022zr,daw2021sluggish}. However, this methodology exhibits inherent limitations, as atomic radii are not fixed values but rather variable parameters influenced by crystal structure and the electronic environment of neighboring atoms. Advanced theoretical modeling, particularly first-principles calculations, have been employed to assess lattice distortions in RHEAs, revealing that local lattice distortions are substantially more pronounced than what the average atomic size mismatch would suggest \cite{lee2018lattice,lee2020lattice}. In particular, element-specific mean squared displacements were calculated and analyzed \cite{sen2023anti,muralikrishna2024microstructure}. In addition, recent developments in experimental techniques utilizing total scattering from synchrotron X-ray or neutron sources have enabled quantitative measurements of the lattice distortions in HEAs. These investigations have demonstrated that such distortions are typically localized within 2--3 unit cells and that the atomic size mismatch represents just one of the multiple factors contributing to lattice distortions in HEAs \cite{tong2018comparison,tong2020severe,tong2018local,thirathipviwat2022role,meng2021charge}.

In our previous study, we conducted measurements of Zr self-diffusion in TiZrHfNbV and TiZrHfNbTa RHEAs, marking the first direct assessment of the self-diffusion rates in BCC HEAs \cite{zhang2022zr}. The findings revealed that Zr self-diffusion in these RHEAs is not sluggish, suggesting that a substantial atomic size mismatch significantly influences diffusion, enhancing its rates. However, a comprehensive understanding of the diffusion mechanisms in BCC HEAs necessitates additional measurements of both diffusion and lattice distortions, particularly concerning impurity diffusion and local distortion phenomena. So far, impurity diffusion in HEAs was measured only for several systems, e.g. for Cu in CoCrFeMnNi \cite{gaertner2018tracer}, Zn in B2-ordered AlCoCrFeNi \cite{Mohan2020-B2}, or for Zn in HfTiZrAlSc  \cite{sen2023does}, leaving a large gap in the current knowledge of diffusion phenomena in multi-principal element alloys. 
Furthermore, specific diffusion mechanisms of impurity atoms in terms of potential interstitial and/or substitutional contributions, as well as the possible impact of short-range order (SRO), are not fully understood.

The present study aims to fill this gap by extending our previous work on Zr diffusion\cite{zhang2022zr} by impurity diffusion measurements on the same batches of TiZrHfNbV and TiZrHfNbTa RHEAs. Diffusion of Co, Zn, and Mn is investigated using the radiotracer technique and applying suitable radioisotopes, $^{57}$Co, $^{65}$Zn, and $^{54}$Mn, respectively.
The selection of Co is motivated by the availability of extensive literature on Co diffusion in the constituent elements of these RHEAs, which facilitates an in-depth analysis of the parameters affecting diffusion. Zn and Mn represent solutes with different valency states that potentially affect the vacancy--solute binding. The lattice distortions in TiZrHfNbV and TiZrHfNbTa RHEAs are quantitatively assessed via a neutron total scattering technique and compared with the results of density-functional-theory (DFT) calculations. Furthermore, the impurity diffusion mechanisms are discussed in relation to substitutional and interstitial solution energies as predicted by the \textit{ab initio} calculations, including a potential impact of SRO in these alloys. This comprehensive investigation explores the intricate relationships between self-diffusion, impurity diffusion, and the degree of lattice distortions in industrially relevant BCC RHEA systems.

\section{Experimental details}
\subsection{Manufacturing of the alloys}
The TiZrHfNbV and TiZrHfNbTa alloys employed in the diffusion measurements were identical to those used in our previous investigation of Zr diffusion \cite{zhang2022zr}. For the neutron scattering experiments, a new batch of specimens was prepared following the same fabrication protocol as previously established. The chemical compositions were checked to be similar for the two batches (see below, Section~\ref{sec:chem}).

High-purity elements (99.99\%) in granular form (2-5 mm) were used for production. Before melting, the granules were ultrasonically cleaned in ethanol. The alloys were arc-melted under an argon atmosphere on a water-cooled copper chill plate. The produced buttons were consolidated into rod-shaped ingots approximately 100 mm in length, with diameters of 6.5 mm for TiZrHfNbTa and 12 mm for TiZrHfNbV. For further details, see Ref.~[\onlinecite{zhang2022zr}].

The rods were sectioned into 25 mm segments and ultrasonically cleaned in ethanol. The samples were then encapsulated in quartz ampoules and annealed at $1200^\circ$C for 48 hours in high vacuum (better than $10^{-5}$~Pa). Subsequently, the samples were water-quenched to room temperature; the cooling rates were estimated to be about 200~K/s.

\subsection{Alloy characterization}
The chemical composition was checked by using an inductively coupled plasma-optical emission spectrometer (ICP-OES, iCAP 7600; Thermo Scientific) to verify the nearly equiatomic composition of both high-entropy alloys. 

Neutron diffraction and total scattering measurements were carried out on the Multi-Physics Instrument (MPI), a total scattering neutron time-of-flight diffractometer at the China Spallation Neutron Source (CSNS)\cite{xu2021multi}. Each sample, approximately 6~g, was put into a vanadium-nickel alloy tank with a diameter of 8.9~mm, followed by placing the can into an automatic sample-changing device. The diffraction patterns were collected within a neutron wavelength range of $0.1 - 4.5$~{\AA}. The detectors were divided into five banks, covering a scattering angle range ($2\theta$) from $12.54^\circ$ to $170.00^\circ$, thus ensuring a complete coverage of a sufficiently large $Q$ range of $0.5 - 30\,\mathrm{\AA}^{-1}$ while maintaining appropriate  resolution in the $Q$ space. A series of standardized data correction procedures was applied to ensure the accuracy and reliability of the resulting structural information. These corrections include incident beam normalization, cross-section normalization based on elemental scattering properties, detector solid angle and efficiency correction (performed using a vanadium standard), application of the Lorch modification function to minimize Fourier transform artifacts, as well as corrections for multiple scattering, background scattering from sample environments, and neutron absorption. Data corrections were performed using the Mantid software \cite{arnold2014mantid}. For further details on the experimental procedures and data processing workflows of the MPI, see e.g. \cite{pan2023long,sun2024reaction,yang2024unveiling,gou2023additive}. Subsequently, the data were Fourier transformed from reciprocal space to real space, resulting in the experimental pair distribution functions (PDFs).

\subsection{Tracer diffusion measurements}
Disc specimens were prepared using electrical discharge machining to dimensions of 1.5 mm in thickness and diameters of 6.5 mm and 12 mm for TiZrHfNbV and TiZrHfNbTa, respectively. Prior to the diffusion experiments, the mechanically polished discs were annealed to relieve the mechanical stresses and establish equilibrium defect concentrations at the target diffusion temperatures. The heat treatments were conducted in evacuated quartz tubes under a high-purity (99.999\%) argon atmosphere.

$^{57}$Co, $^{65}$Zn, and $^{54}$Mn radioisotopes were provided by PerkinElmer LAS GmbH, Germany; for details, see Table~\ref{table:isotopes}. The initial oxalic acid solutions were diluted with double-distilled water and deposited onto the mirror-polished surfaces of the disc specimens. After drying, diffusion annealing was performed in sealed quartz tubes under a high-purity (99.999\%) argon atmosphere at the temperatures and durations specified in Table~\ref{tab:CoMnZndata}. The furnace temperature was controlled within $\pm1$~K using a certified Ni/NiCr thermocouple (type K).

After annealing, the penetration profiles were determined using serial sectioning via high-precision parallel grinding. Sections ranging from 0.3 to 10 $\mu$m were ground. The section thicknesses were determined with a relative accuracy of better than $\pm 0.03$\% by meticulously weighing the samples before and after the grinding steps using a high-precision microbalance (about $\pm 0.1$~{\textmu}g). The intensity of the $\gamma$-decays was determined by a solid Ge-detector equipped with a multi-channel 16 K energy discriminator. Proper counting times were chosen to provide statistical measurement uncertainties of better than 2\% for all relevant $\gamma$-energy windows.

\begin{table}[ht]
\centering
\caption{Decay modes of the applied radioisotopes. The characteristic $\gamma$ energies are listed.}
\label{table:isotopes}

\begin{tabular}[t]{cccc}

\hline
\multirow{2}{*}{Radioisotopes} & Half-life & \multirow{2}{*}{Decay mode} & $\gamma$ energy\\
 & (day) & & (keV)\\
\hline
$^{57}$Co&271.8&$\epsilon$&14.4, 122.1 \& 136.5\\
$^{54}$Mn&312.2&$\epsilon$, $\beta^{+}$ \& $\beta^{-}$&834.9\\
$^{65}$Zn&244.0&$\epsilon$ \& $\beta^{+}$&345.0, 770.6 \& 1115.6\\
\hline
\end{tabular}
\end{table}%

\begin{table}[ht]
\renewcommand{\arraystretch}{0.85}
\centering
\caption{Annealing temperatures $T$, times $t$, and the determined diffusion coefficients $D$ for $^{57}$Co, $^{54}$Mn, and $^{65}$Zn in the TiZrHfNbV and TiZrHfNbTa RHEAs. The typical experimental uncertainties of the $D$ values are estimated to be less than about 20\%. }\label{tab:CoMnZndata}

\begin{tabular}{c|cccc}

\hline
Tracer & Alloy & $T$ (K) & $t$ ($10^{3}$ s) & $D$ (m$^2$/s) \T\B \\
\hline
\multirow{8}{*}{$^{57}$Co} & \multirow{4}{*}{TiZrHfNbV} & 1073 & 79.2 & $7.80 \times 10^{-15}$ \T \\
  & & 1123 & 12.6 & $2.16 \times 10^{-14}$ \\
  & & 1173 & 7.2 & $5.63 \times 10^{-14}$ \\
  & & 1223 & 1.8 & $1.54 \times 10^{-13}$ \\
\cline{2-5}
  & \multirow{4}{*}{TiZrHfNbTa} 
  & 1073 & 79.2 & $1.46 \times 10^{-14}$ \\
  & & 1123 & 12.6 & $3.10 \times 10^{-14}$ \\
  & & 1173 & 7.2 & $6.15 \times 10^{-14}$ \\
  & & 1223 & 1.8 & $1.34 \times 10^{-13}$ \B \\
\hline
                          \multirow{12}{*}{$^{54}$Mn} & \multirow{5}{*}{TiZrHfNbV} & 1073 & 79.2 & $1.00 \times 10^{-15}$ \\
                           & & 1173 & 7.2 & $9.84 \times 10^{-15}$ \\
                            & & 1223 & 172.8 & $1.28 \times 10^{-14}$ \\
                            & & 1273 & 57.6 & $3.97 \times 10^{-14}$ \\
                            & & 1323 & 43.2 & $9.85 \times 10^{-14}$ \\
                           \cline{2-5}
                          & \multirow{7}{*}{TiZrHfNbTa} & 1073 & 79.2 & $1.18 \times 10^{-15}$ \\
                           & & 1123 & 12.6 & $3.10 \times 10^{-15}$ \\
                           & & 1123 & 923.4 & $3.42 \times 10^{-15}$ \\
                           & & 1173 & 7.2 & $6.00 \times 10^{-15}$ \\
                           & & 1223 & 1.8 & $1.17 \times 10^{-14}$ \\
                           & & 1223 & 172.8 & $1.07 \times 10^{-14}$ \\
                          & & 1273 & 57.6 & $1.86 \times 10^{-14}$ \\
                           \hline
                          \multirow{11}{*}{$^{65}$Zn} & \multirow{5}{*}{TiZrHfNbV} & 1073 & 172.8 & $1.24 \times 10^{-15}$ \T \\
                           & & 1173 & 57.6 & $7.65 \times 10^{-15}$ \\
                           & & 1223 & 43.2 & $1.04 \times 10^{-14}$ \\
                           & & 1273 & 43.2 & $2.42 \times 10^{-14}$ \\
                          & & 1323 & 5.4 & $4.93 \times 10^{-14}$ \B \\
\cline{2-5}
& \multirow{6}{*}{TiZrHfNbTa} & 1073 & 79.2 & $1.89 \times 10^{-16}$ \\
                           & & 1123 & 923.4 & $6.95 \times 10^{-16}$ \\
                            & & 1173 & 172.8 & $6.48 \times 10^{-16}$ \\
                            & & 1223 & 7.8 & $1.57 \times 10^{-15}$\\
                            & & 1223 & 172.8 & $2.41 \times 10^{-15}$\\
                            & & 1273 & 57.6 & $2.88 \times 10^{-15}$\\
                             \hline
\end{tabular}
\end{table}

\section{Computational details}

\subsection{Density functional theory calculations}

For the \textit{ab initio} calculations of lattice distortions and solution energies in TiZrNbHfTa and TiZrNbHfV, we used DFT in combination with special quasirandom structures (SQSs)~\cite{zunger1990special} to effectively mimic the random single-phase solid solutions. Most calculations were performed utilizing a supercell containing 128 atoms (4$\times$4$\times$4 expansion of the BCC unit cell), consistent with our previous study~\cite{vaidya2020phenomenon}. For local distortion calculations, test calculations with larger 250-atom SQS cells (5$\times$5$\times$5) were performed, too. A possible impact of SRO (not included in the SQS approach) and supercell size effects were analyzed using a foundation model and Monte Carlo simulations (technical details in Sec.~\ref{Sec:Method_SRO}).

For the DFT calculations, the projector augmented wave method~\cite{blochl1994projector}, as implemented in VASP~\cite{kresse1996efficiency,kresse1996efficient}, was used with the generalized gradient approximation (GGA) in the parametrization of Perdew, Burke, and Ernzerhof (PBE)~\cite{perdew1996generalized}. 
The plane wave cut-off was set to 230 eV for all calculations. 
The Methfessel–Paxton scheme~\cite{methfessel1989high} of order 1 with a smearing value of 0.1 eV was used to treat orbital partial occupancies. 
Integrations in reciprocal space were based on a $\Gamma$-centered Monkhorst-Pack grid~\cite{monkhorst1976special} with a mesh of $2\times 2 \times 2$. Non-spin-polarized calculations were conducted for the RHEAs, except for determining the reference energies of elemental Co and Mn.

\subsection{Radial distribution function and element-specific lattice distortions}
To decode the lattice distortions revealed by the neutron PDFs, we computed radial distribution functions and element-specific lattice distortions through the following protocol. Initially, the atomic positions of the host matrices (without impurity atoms) were relaxed while keeping the shape and volume of the supercell fixed, until the atomic forces converged to below 0.05 eV/\AA. 
The equilibrium volume for each alloy matrix was obtained by fitting a dense energy-volume dataset (11 volumes ranging from 3.10 to 3.60 \AA) with the Vinet equation of state and was utilized for further calculations of the structures with the impurity atoms, i.e., those with substitutional and interstitial Co, Mn, and Zn.

The radial distribution functions were obtained by examining the distribution of the distances between all atomic pairs in the relaxed structure, using the OVITO software~\cite{stukowski2009visualization}.
For each alloying element the displacements from the relaxation were used to calculate the mean-squared displacement (MSD),
\begin{equation}
    {\rm MSD} = \frac{1}{N} \sum_{i} \| \vec{R}_i - \vec{R}_{i,0} \|^2,
    \label{eq:msd}
\end{equation}
where $\vec{R}_i$ and $\vec{R}_{i,0}$ represent the relaxed and ideal atomic coordinates of atom $i$ of a given alloying element and $N$ the number of atoms of that element in the alloy. The PDF, \( G(r) \), derived from the radial distribution function, \( g(r) \), was obtained as\cite{keen2001comparison}
\begin{equation}\label{eq:PDFRDF}
G(r) = 4\pi r \rho_0 \left[ g(r) - 1 \right],
\end{equation}
where \( \rho_0 \) is the average atomic number density of the system and $r$ is the interatomic distance.

\subsection{Solution energies of impurity elements}
An accurate determination of impurity diffusion coefficients in HEAs, e.g., by \textit{ab initio} molecular dynamics presents a tremendous task. Even classical kinetic Monte Carlo calculations with a list of barriers or barriers calculated on-the-fly are time-consuming in view of the chemical complexity and the required statistics. In order to provide \textit{ab initio} insights into potential diffusion mechanisms, we computed with DFT the solution energies of Co, Mn, and Zn in both substitutional (sub) and interstitial (int) positions, following the scheme from Ref.~[\onlinecite{sen2023anti}]. 
The formation energies $\Delta E_p^{\rm sub}$ and $\Delta E_p^{\rm int}$ of the substitutional and interstitial impurities were extracted based on the calculated total energies $E$ and chemical potentials $\mu$ by
\begin{equation}\label{eq:3}
   \Delta E_p^{\rm sub} = E\left[ {\rm SQS}-X_p+ {\rm Y}^{\rm sub}_p \right] - E\left[ {\rm SQS} \right] + \mu_{{\rm X}_p} - \mu_{\rm Y},
\end{equation}
\begin{equation}\label{eq:4}
    \Delta E_p^{\rm int} = E\left[ {\rm SQS}+ {\rm Y}^{\rm int}_p \right] - E\left[ {\rm SQS} \right] - \mu_{\rm Y},
\end{equation}
where SQS means the defect-free SQS-generated structure, ${\rm Y}=$ Co, Mn, or Zn depending on the impurity under consideration, and $\mu_{\rm Y}$ is the corresponding chemical potential of the impurities. The subscript $p$ indexes the substitutional atomic site being replaced and the associated interstitial positions, with $\mu_{{\rm X}_p}$ denoting the chemical potential of the replaced element (${\rm X}_p$ = Ti, Zr, Nb, Hf, and Ta/V). The chemical potentials of Co and Zn were calculated using an HCP unit cell, whereas the $\alpha$-Mn (A12) structure was used for Mn.
Spin-polarization was considered for pure Co and Mn.

A single Co, Mn, or Zn impurity atom representing the dilute limit was introduced into the perfect SQS supercell as follows. For the substitutional case, every site in the SQS supercell was once replaced by a Co/Mn/Zn atom, yielding 128 defect supercells with a single substitutional Co/Mn/Zn. For our interstitial calculations, we focused on octahedral sites. Test calculations on approximately 100 tetrahedral sites for Co indicated that both tetrahedral and octahedral configurations relax into similar dumbbell-like structures. We sampled all 384 octahedral positions in both host alloys for Co to assess statistical convergence. Based on this analysis, around 100 sites per host alloy were utilized for Mn and Zn.

The total energies of structures with a defect (substitutional or interstitial Co, red{Mn}, or Zn) were obtained as follows. 
The reference perfect SQS structures were firstly fully relaxed using the conjugate-gradient algorithm. 
One type of defect was then introduced to the relaxed perfect SQS configurations followed by an ionic relaxation only (cell shape and volume fixed). 
The relaxation was stopped when the atomic forces converged to below 0.05 eV/\AA. 

\subsection{Short-range order calculations}\label{Sec:Method_SRO}

Short-range order was investigated using on-lattice low-rank potentials (LRP) \cite{shapeev2017,meshkov2019}, which were trained on calculations using the
semi-local, non-linear Graph Atomic Cluster Expansion (GRACE-2L) framework \cite{PhysRevX.14.021036} and utilizing the GRACE-2L-OMAT foundation model based on the Open Materials (OMAT) database \cite{gracemaker_foundation}. The GRACE calculations were benchmarked against DFT solution energies and equations of state for the 128-atom supercells. All calculations were performed relaxing the atomic positions with a force convergence criterion of 0.01 eV/\AA. 

The LRP training and subsequent solution-energy calculations made use of larger 250-atom cells with concentration ratios for the alloys, specifically Hf–Nb–Ti–V–Zr (50 : 56 : 47 : 49 : 48 atoms) and Hf–Nb–Ta–Ti–Zr (58 : 56 : 40 : 48 : 48 atoms). The initial training set comprised 500 random configurations per alloy. Following a preliminary Monte Carlo (MC) calculation from 500 to 2000 K, we added 480 additional configurations selected from MC snapshots between 1340 and 400 K (10 snapshots at each temperature in 20 K steps). For all calculations, LRPs of rank 5 were selected, resulting in training and validation errors below 2 meV per atom (using 90$\%$ of configurations for training and 10$\%$ for validation).

To account for the effect of thermal expansion, we utilized the quasi-harmonic Debye model, evaluated at 1000 K. This resulted in an increased volume of 2.76$\%$ for TiZrNbHfTa and 3.80$\%$ for TiZrNbHfV.

The LRP training and the subsequent MC simulations were conducted at these larger volumes. The MC sampling was carried out in 250-atom cells with periodic boundary conditions for 40,000 steps per atom, discarding the first half for equilibration. Test runs in 6750-atom ($15 \times 15 \times 15$) cells confirmed that the supercell size had only a minor effect on the SRO parameters at 1000~K. To quantify the impact of ordering, three equilibrated 250-atom snapshots at 1000~K were selected for the analysis of solution energies and were compared with three fully random supercells.

\color{black}

\section{Results}

\subsection{Microstructure analysis}\label{sec:chem}

The compositions of the two BCC RHEAs used for the present neutron scattering experiments were determined with ICP-OES and obtained as   Ti$_{19.2}$Zr$_{19.3}$Nb$_{22.2}$Hf$_{23.0}$Ta$_{16.3}$ and Ti$_{18.6}$Zr$_{19.0}$Nb$_{22.2}$Hf$_{20.1}$V$_{20.1}$  (in at.\%). These compositions are very similar to the alloys used for the radiotracer experiments; the compositions of the latter were measured with micro-X-ray fluorescence analysis and reported as Ti$_{17.9}$Zr$_{20.1}$Nb$_{22.0}$Hf$_{20.3}$Ta$_{19.7}$ and Ti$_{19.7}$Zr$_{19.9}$Nb$_{20.7}$Hf$_{19.2}$V$_{20.5}$ (in at.\%)\cite{zhang2022zr}. 

Previous characterization\cite{zhang2022zr} through X-ray diffraction (XRD), energy-dispersive X-ray spectroscopy (EDS), and electron back-scatter diffraction (EBSD) confirmed several critical microstructure features: a single-phase (solid solution) BCC structure, grain sizes exceeding $100$~{\textmu}m, uniform elemental distribution, and the absence of preferred crystallographic orientations (texture). The lattice parameters were measured as 0.3374~nm for TiZrHfNbV and 0.3403~nm for TiZrHfNbTa, respectively. 
The DFT-predicted lattice parameters are 0.3347\,nm for TiZrHfNbV and 0.3403\,nm for TiZrHfNbTa. 

Further details on the coarse-grained microstructures of the two high-entropy alloys after appropriate homogenization treatment are given in Ref.~[\onlinecite{zhang2022zr}]. The microstructures remain largely unchanged after the heat treatments used for the present diffusion measurements. In TiZrHfNbV, a relatively strong V segregation to grain boundaries was seen after annealing treatment below 1323 K, which, however, does not significantly influence the volume diffusion experiments.  According to the exact solution of the grain boundary diffusion problem \cite{Aloke2014}, the tracer concentration at near-surface depths below approximately $2\sqrt{Dt}$ is dominated by direct volume diffusion from the sample surface, where $D$ denotes the volume diffusion coefficient (specifically the Zr self-diffusion coefficient or the Co, Zn, and Mn impurity diffusion coefficients in this study) and $t$ represents the diffusion time. The total amount of V atoms segregated at grain boundaries is negligibly small and does not affect strongly the alloy composition in the present case in view of a relatively large grain size in the coarse-grained microstructure, i.e., because of a small total volume fraction of the grain boundary-related material (which can be estimated to be less than $10^{-5}$ assuming the grain boundary width of about 1 nm) and a relatively small grain boundary excess of V atoms.

\subsection{Diffusion measurements}

\begin{figure}
	\centering
		\includegraphics[width=0.495\textwidth]{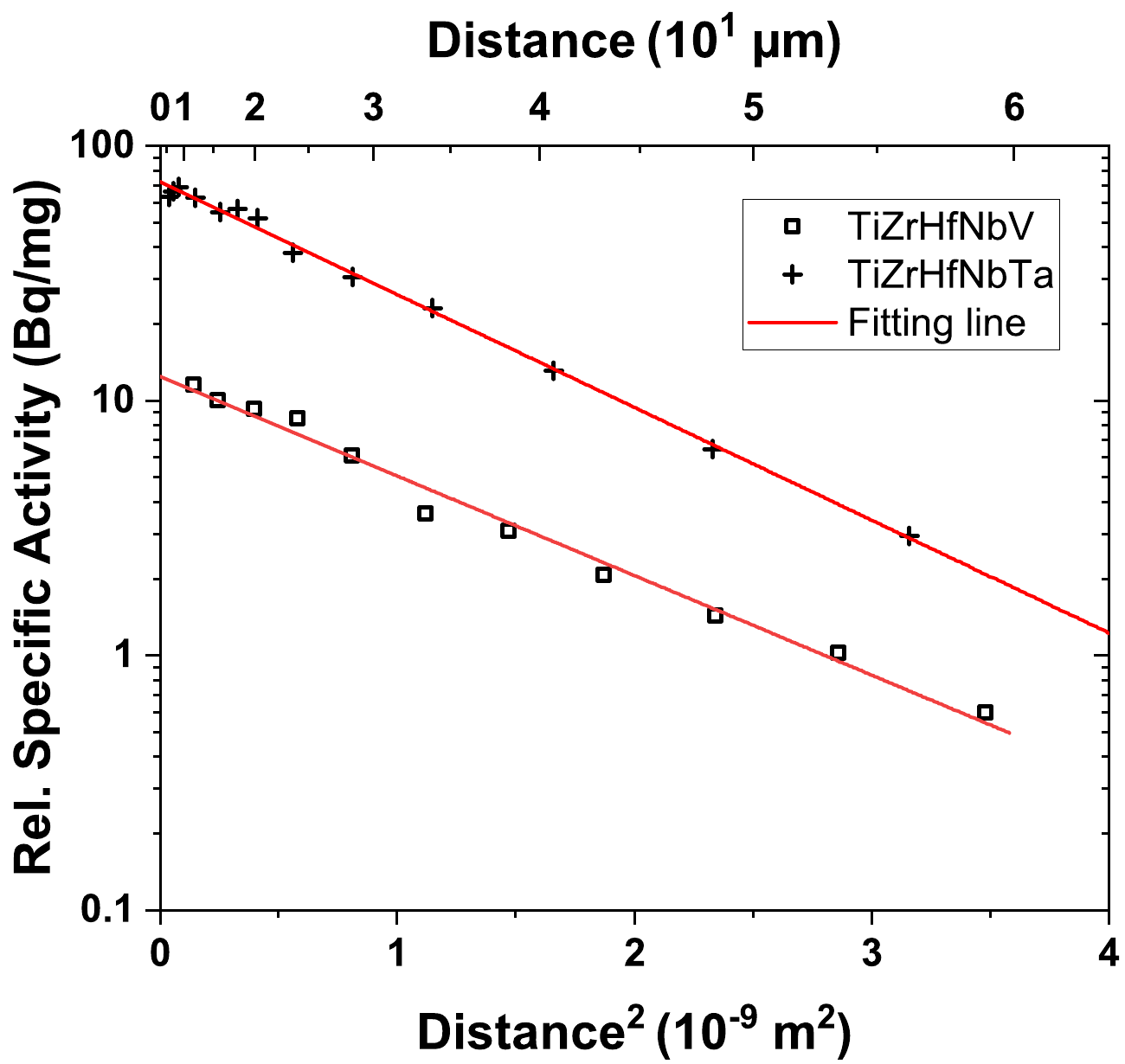}
	\caption{Concentration profiles of Co tracer diffusion measured at 1223 K in the TiZrHfNbV and TiZrHfNbTa RHEAs. The red solid lines represent fits according to bulk diffusion from an instantaneous source, Eq.~(\ref{eq:Gauss}).}
	\label{fig:profiles}
\end{figure}

As an example, the concentration profiles of Co diffusion in the TiZrHfNbV and TiZrHfNbTa RHEAs measured at 1223~K are shown in Fig.~\ref{fig:profiles}. Other penetration profiles are of a similarly high quality. 
 
\begin{table}[h!]
\centering
\caption{Pre-exponential factors $D_0$ and activation enthalpies $Q$ for bulk diffusion of $^{57}$Co, $^{54}$Mn, and $^{65}$Zn (determined in this work) as well as Zr\cite{zhang2022zr} in the  TiZrHfNbV and TiZrHfNbTa RHEAs.}\label{tab:Arrhenius}
\begin{tabular}{l | c c c c}
\hline
Tracer~~~~ & Alloy & ~~~~~~~$D_0$ (m$^2$/s)~~~ & ~~~$Q$ (kJ/mol)~~~\\
\hline
\multirow{2}{*}{$^{57}$Co} & TiZrHfNbV & $(2.5^{+2.1}_{-1.1}) \cdot 10^{-4}$ & $215.4 \pm 7.9$ \\
 & TiZrHfNbTa & $(8.3^{+6.9}_{-3.8}) \cdot 10^{-7}$ & $159.6 \pm 5.7$ \\
\hline
\multirow{2}{*}{$^{65}$Zn} & TiZrHfNbV & $(2.2^{+4.1}_{-1.4}) \cdot 10^{-7}$ & $169.3 \pm 10.5$ \\
 & TiZrHfNbTa & $(1.5^{+7.0}_{-1.3}) \cdot 10^{-6}$ & $202.4 \pm 16.3$ \\
 \hline
\multirow{2}{*}{$^{54}$Mn} & TiZrHfNbV & $(1.62^{+6.7}_{-1.3}) \cdot 10^{-5}$ & $209.6 \pm 16.3$ \\
 & TiZrHfNbTa & $(3.4^{+3.7}_{-1.8}) \cdot 10^{-8}$ & $151.9 \pm 5.7$ \\
\hline
\multirow{2}{*}{Zr\cite{zhang2022zr}} & TiZrHfNbV & $4.8 \cdot 10^{-9}$ & $134.2$ \\
 & TiZrHfNbTa & $3.8 \cdot 10^{-8}$ & $172.4$ \\
\hline
\end{tabular}
\end{table}

The instantaneous source (Gaussian) solution of the bulk diffusion problem was found to be appropriate under the present conditions, Fig.~\ref{fig:profiles}, and the corresponding branches of the concentration profiles were approximated as\cite{paul2014thermodynamics}
\begin{equation}\label{eq:Gauss}
C\left(x,t\right)=\frac{M}{\sqrt{\pi Dt}}\exp{\left(-\frac{x^2}{4Dt}\right)}.
\end{equation}
Here, $C\left(x,t\right)$ is the concentration at a depth $x$ from the surface after the time $t$, $D$ is the determined volume diffusion coefficient, and $M$ is the initial tracer amount.

The determined volume diffusion coefficients of $^{57}$Co, $^{54}$Mn, and $^{65}$Zn are reported in Table~\ref{tab:CoMnZndata} together with the experimental conditions for the two RHEAs under investigation. In a single measurement, the estimated uncertainty of the determined tracer diffusion coefficients is below 20\% and corresponds mainly to the temperature uncertainties.

The temperature dependencies of the tracer diffusion coefficients were found to follow reasonably well Arrhenius-type dependencies,

\begin{equation}\label{eq:Arrhenius}
D=D_0 \exp{\left(-\frac{Q}{RT}\right)},
\end{equation}
with the pre-exponential factor $D_0$ and the activation enthalpy $Q$. The determined pre-exponential factors and activation enthalpies for bulk diffusion of $^{57}$Co, $^{65}$Zn, and $^{54}$Mn in the two RHEAs under consideration are summarized in Table~\ref{tab:Arrhenius}.

\begin{figure*}[h]
\centering 
\includegraphics[width=0.495\textwidth]{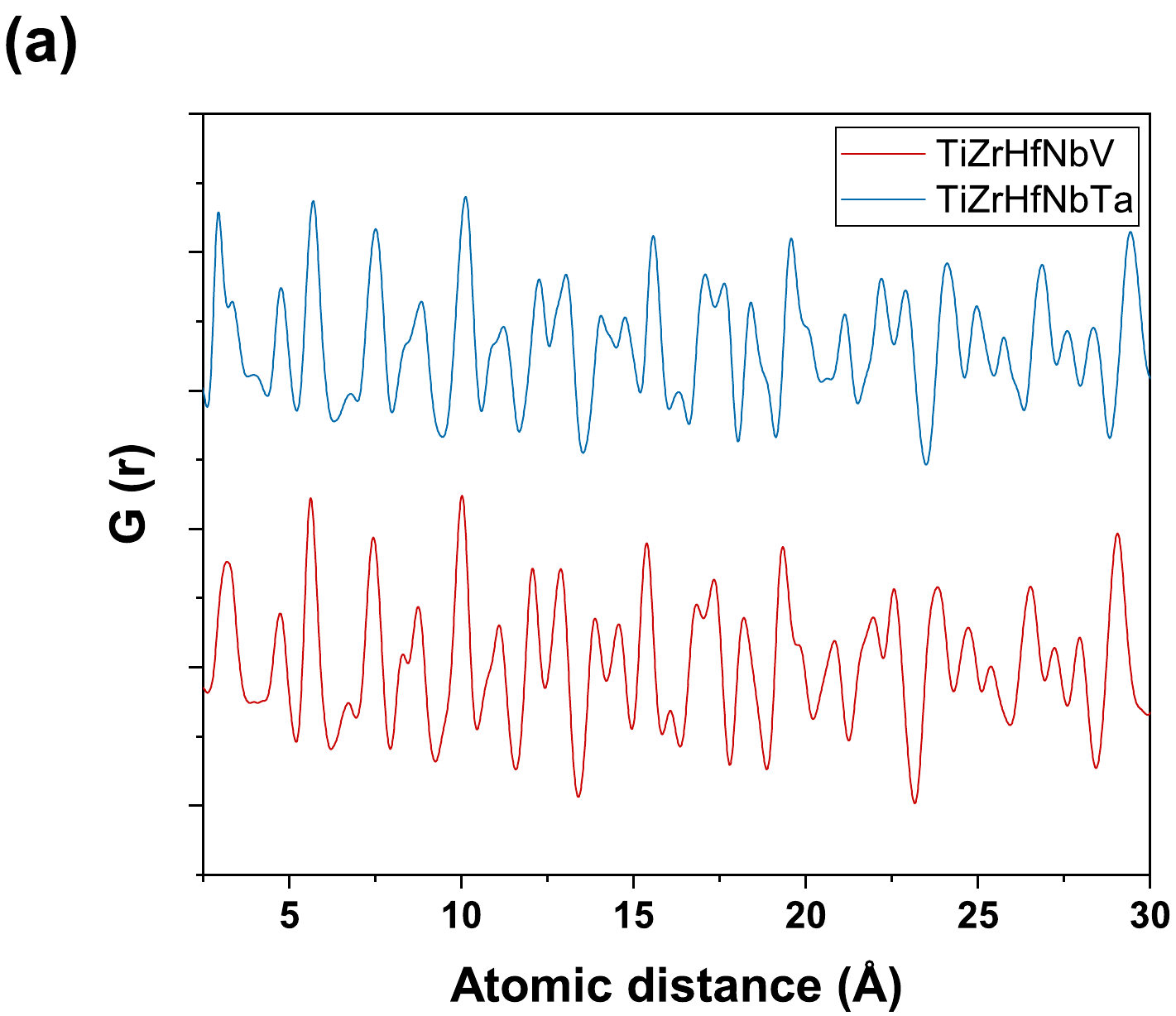}
\hfill
\includegraphics[width=0.495\textwidth]{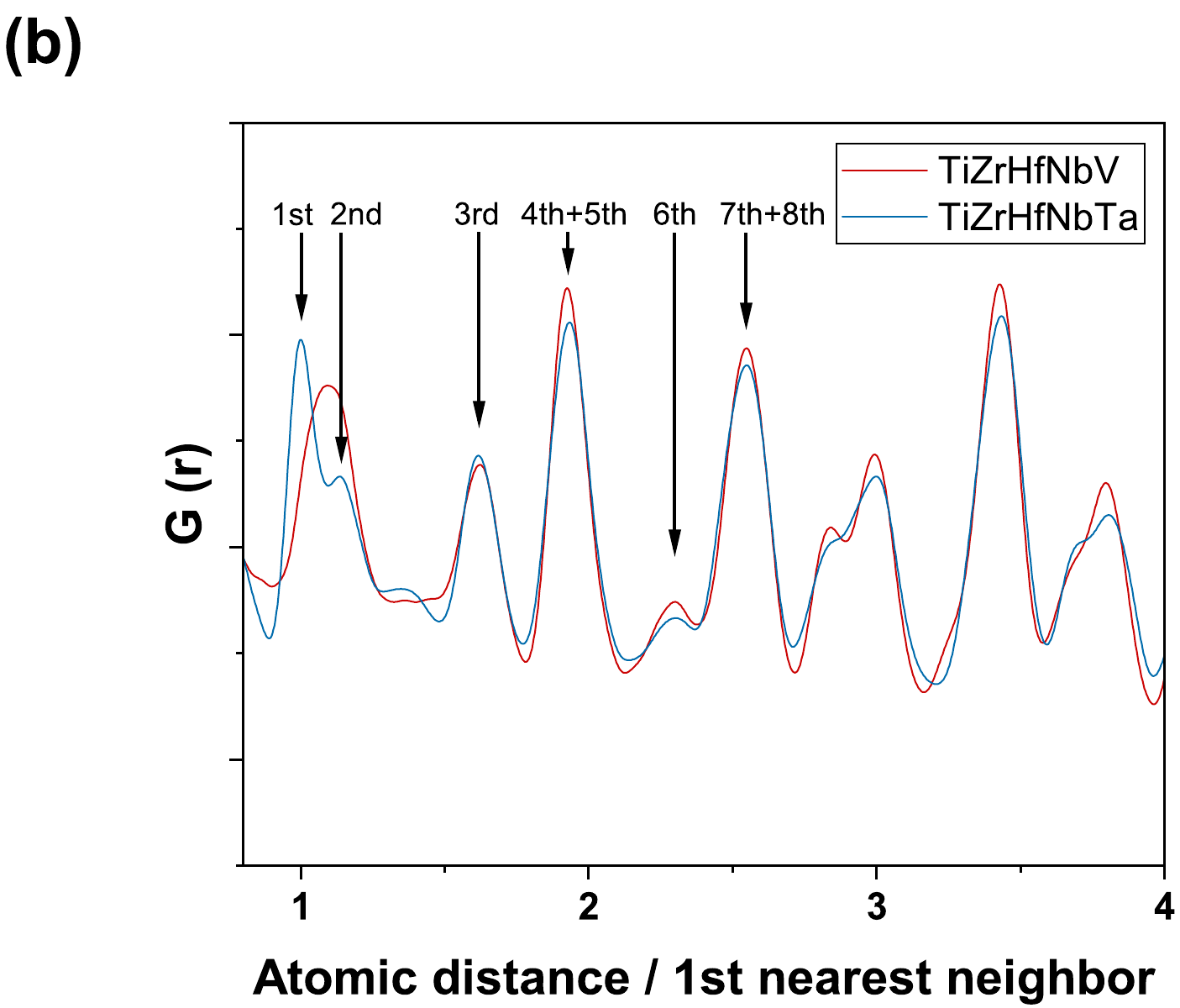}
\hfill
\includegraphics[width=0.495\textwidth]{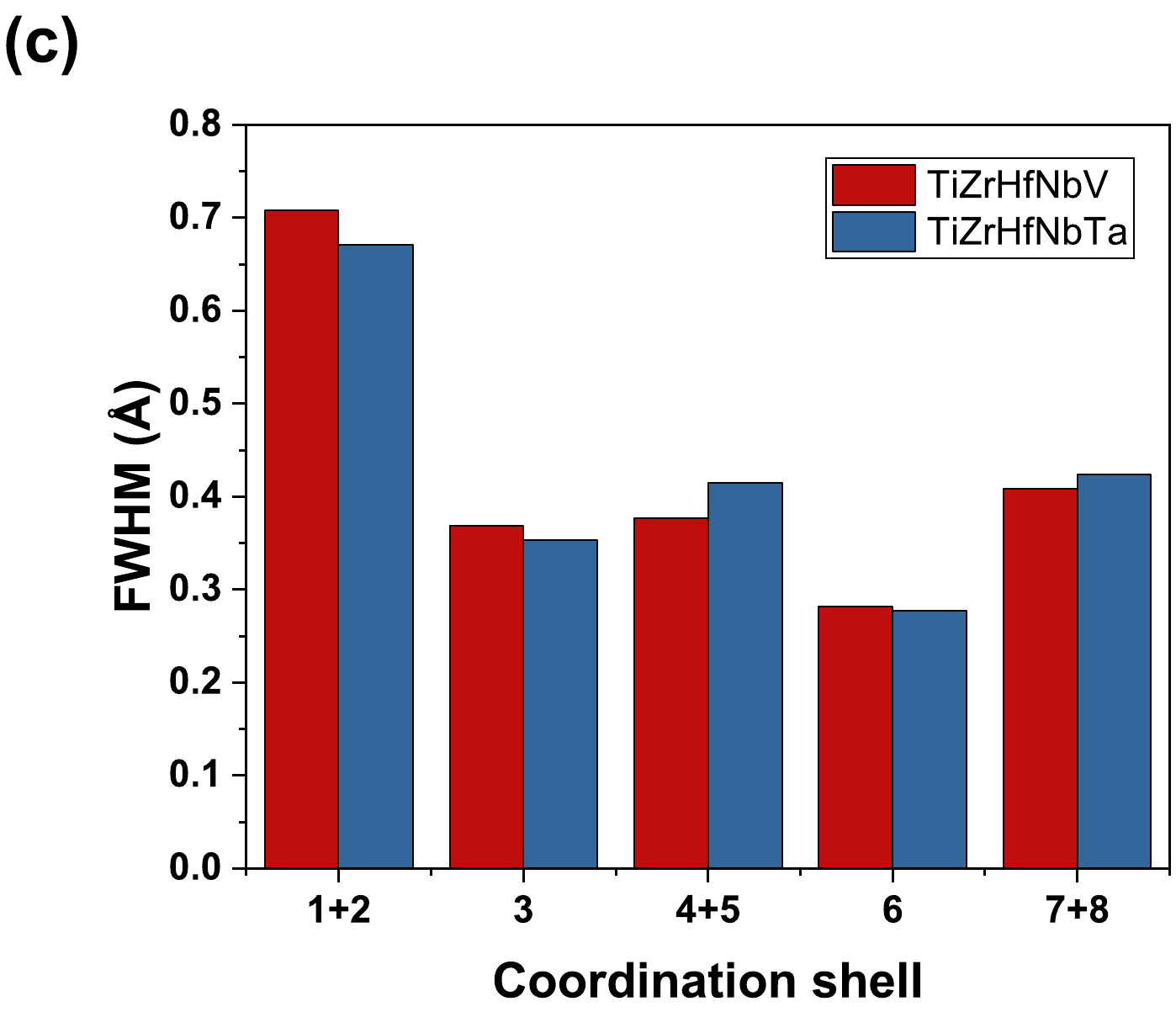}
\hfill
\includegraphics[width=0.495\textwidth]{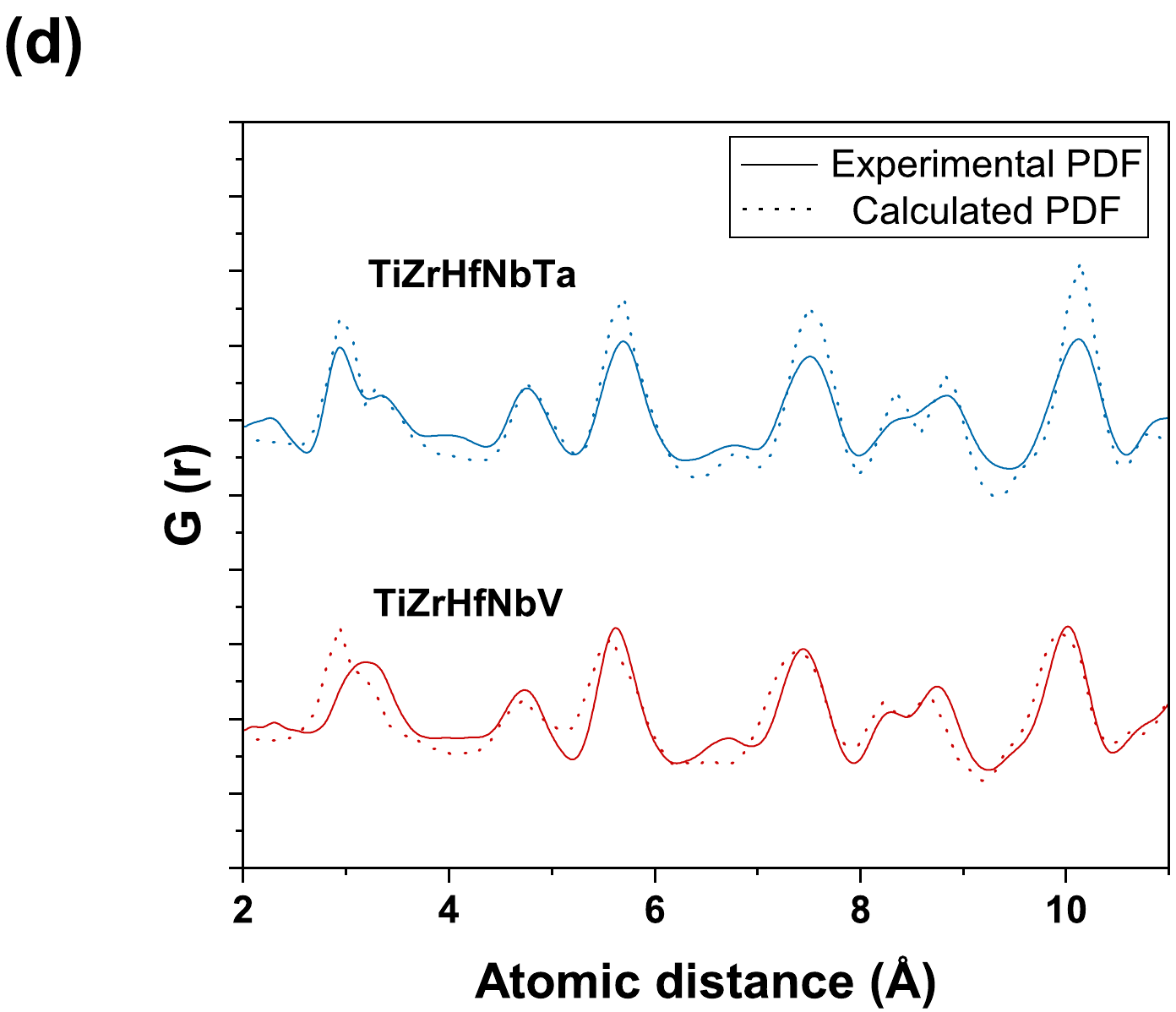}

\caption{(a) PDF profiles, (b) PDF profiles for the first 8 peaks normalized by the 1st nearest neighbor, (c) the FWHM of the first 8 coordination shells, and (d) a comparison of the PDFs obtained from neutron total scattering experiments and those derived from radial distribution functions obtained through DFT calculations for the TiZrHfNbV and TiZrHfNbTa RHEAs.}
\label{fig:PDF}
\end{figure*} 

\subsection{Lattice distortions}
To investigate atomic-scale structural variations in the two RHEAs, neutron total scattering experiments were conducted. The position and shape of the peaks in a PDF reveal information about the local atomic structure at a given interatomic distance $r$. In a distorted lattice, the peak widths in the PDF increase as atoms deviate from their ideal positions. The PDFs for both RHEAs are shown in Fig.~\ref{fig:PDF}(a). The analysis is focused on the first 8 peaks in Fig.~\ref{fig:PDF}(b), particularly the first peak, due to the stronger bonding interactions between nearest neighbor atoms. Each peak was fitted with a Gaussian function to determine the full width at half maximum (FWHM), see Fig.~\ref{fig:PDF}(c). 

In Fig.~\ref{fig:PDF}(d), the PDFs obtained from neutron total scattering experiments and those derived from radial distribution functions obtained through DFT calculations for the TiZrHfNbV and TiZrHfNbTa RHEAs are compared. The variations in peak separation observed in the experimental PDFs for both RHEAs are accurately reproduced in the calculated PDFs, highlighting a strong agreement between the experimental and computational results.

\section{Discussion}

\subsection{Lattice distortions in the TiZrHfNbV and TiZrHfNbTa RHEAs}

The conventional approach to evaluating lattice distortions accounts for the atomic size mismatch $\delta$ based on atomic radii as\cite{zhang2022zr,daw2021sluggish}

\begin{equation}\label{eq:delta}
\delta=\sqrt{\sum_{i=1}^{n} x_{i}\left(1-\frac{R_{i}}{\sum_i x_{i} R_{i}}\right)^{2}},
\end{equation}
where $x_{i}$ is the composition (mole fraction) of the constituent element $i$ and $R_{i}$ is its atomic radius. 

The atomic size mismatch $\delta$ for TiZrHfNbTa and TiZrHfNbV is equal to 3.8\% and 5.9\%, respectively\cite{zhang2022zr}. TiZrHfNbV exhibits larger distortions than TiZrHfNbTa due to the smaller atomic radius of V compared to Ta, creating stronger local strain fields. The delocalized electrons from Ta's 5$d$ orbitals facilitate a more homogeneous bonding with neighboring atoms, resulting in a more uniform interatomic spacing. In contrast, TiZrHfNbV exhibits larger local lattice distortions due to the more localized V 3$d$ electrons and weaker metallic bonds. The combination of a smaller atomic radius of V and more localized electronic states leads to more pronounced local structural variations and a higher lattice strain. Thus, both the atomic radius difference and the electronic environment difference result in larger lattice distortions of TiZrHfNbV compared to TiZrHfNbTa. However, this analysis has inherent limitations, since atomic radii are not element-specific values but rather varying, depending on the crystal structure and the electronic environment. 

As shown in Fig.~\ref{fig:PDF}(a), the PDF obtained through Fourier transformation of the total scattering data reveals the local atomic structures of TiZrHfNbV and TiZrHfNbTa in terms of the averaged interatomic distances. To highlight the distortions induced by substituting Ta with V in the two RHEAs, the PDF profiles of the first eight peaks were normalized to the first nearest neighbor distance, Fig. \ref{fig:PDF}(b). The TiZrHfNbTa RHEA exhibits separated first and second PDF peaks, corresponding to the first and second atomic shells, respectively. In contrast, the TiZrHfNbV RHEA displays an unusual structural feature characterized by the strongly overlapping first and second PDF peaks, indicating significant local lattice distortions. The larger local lattice distortions in the TiZrHfNbV RHEA are further substantiated by the shift of the first PDF peak position. The first PDF peak position (3.195 \AA; Fig. \ref{fig:PDF}(b)) exceeds the value anticipated from the average crystal structure (2.922 \AA~and estimated as $\sqrt{3}a/2$ with $a$ being the lattice constant), suggesting the presence of internal strains within the first atomic shell of the TiZrHfNbV RHEA.

Further quantitative characterization of the PDF peaks can be performed by fitting them with the non-normalized version of the Gaussian function\cite{OWEN2020428,gou2023additive,thirathipviwat2022role}. A distorted lattice increases the full width at the half maximum (FWHM) of each peak in the PDF as the atoms are displaced from their ideal positions. FWHMs of the first 8 peaks of the TiZrHfNbV and TiZrHfNbTa RHEAs are compared in Fig. \ref{fig:PDF}(c). 
 
For the 1st, 2nd, and 3rd coordination shells, the FWHM of TiZrHfNbV RHEA is larger. The first three peaks are more critical for defining the local atomic environment due to stronger bonding interactions between the 
nearest neighbor atoms and lower noise of the Fourier transformation of the structure function\cite{thirathipviwat2022role}. As a result, a larger local lattice distortion in the TiZrHfNbV RHEA is further substantiated by the larger FWHM of the first 3 peaks in the PDF. For the less critical 4th to 8th coordination shells, the FWHM of TiZrHfNbV is generally smaller, except for the 6th peak. The difference in the peak widths between the two RHEAs decreases as the atomic distance increases. 

\begin{figure}[ht]
\centering 
\includegraphics[width=0.9\linewidth]{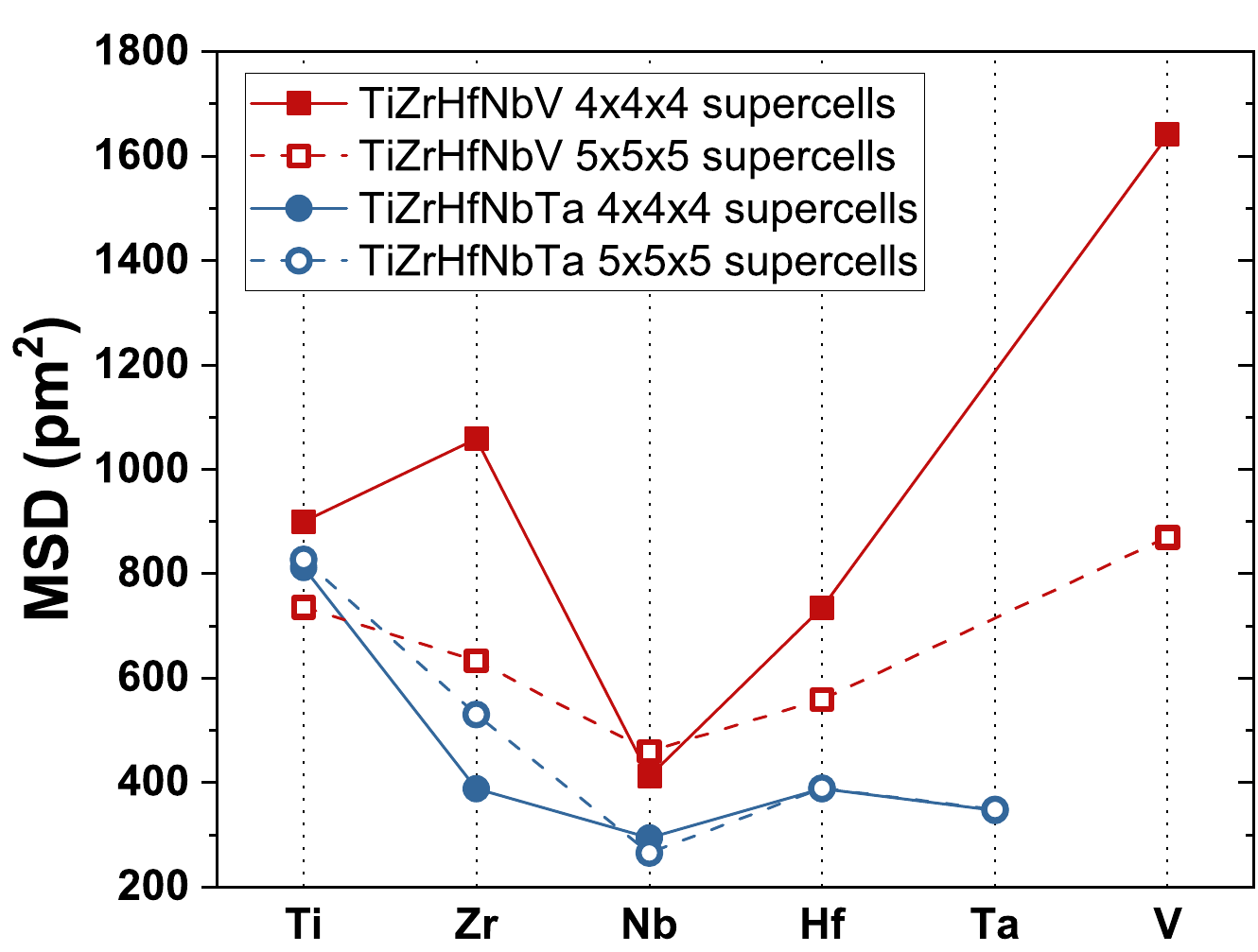}
\caption{Chemically resolved mean squared displacements (MSDs) for different matrix elements from DFT calculations.}
\label{fig:MSD}
\end{figure}

The more substantial local lattice distortions in the TiZrHfNbV RHEA are evident from the DFT-calculated chemically resolved MSDs using Eq.~\eqref{fig:MSD}. Alloying elements in TiZrHfNbV exhibit overall larger MSDs than in TiZrHfNbTa, with V atoms showing the largest MSD among all constituents. This trend persists when larger supercells are used. In the 128-atom supercell, the seemingly large MSDs for V, Zr, and Hf were traced back to two V atoms exhibiting significant relaxations, which cannot be attributed to BCC positions anymore. While this affects the MSD calculations referenced to ideal BCC lattice sites, the impact of the supercell size on solution energies remains small, see Sec.~\ref{sec:volume_SRO}, due to effective cancellation in Eqs.~\eqref{eq:3} and \eqref{eq:4}.

Compared to TiZrHfNbTa, the TiZrHfNbV RHEA exhibits both larger overall lattice distortion calculated through atomic size mismatch and local lattice distortion characterized by PDFs and MSDs. This larger distortion primarily stems from the relatively small atomic size and localized electronic structure of V. In addition, both RHEAs exhibit larger local lattice distortions than conventional refractory alloys or pure metals, which display completely separated 1st and 2nd PDF peaks\cite{thirathipviwat2022role}.

\subsection{Impurity and self-diffusion in the TiZrHfNbV and TiZrHfNbTa RHEAs}

\begin{figure*}[ht]
\centering 
\includegraphics[width=0.495\textwidth]{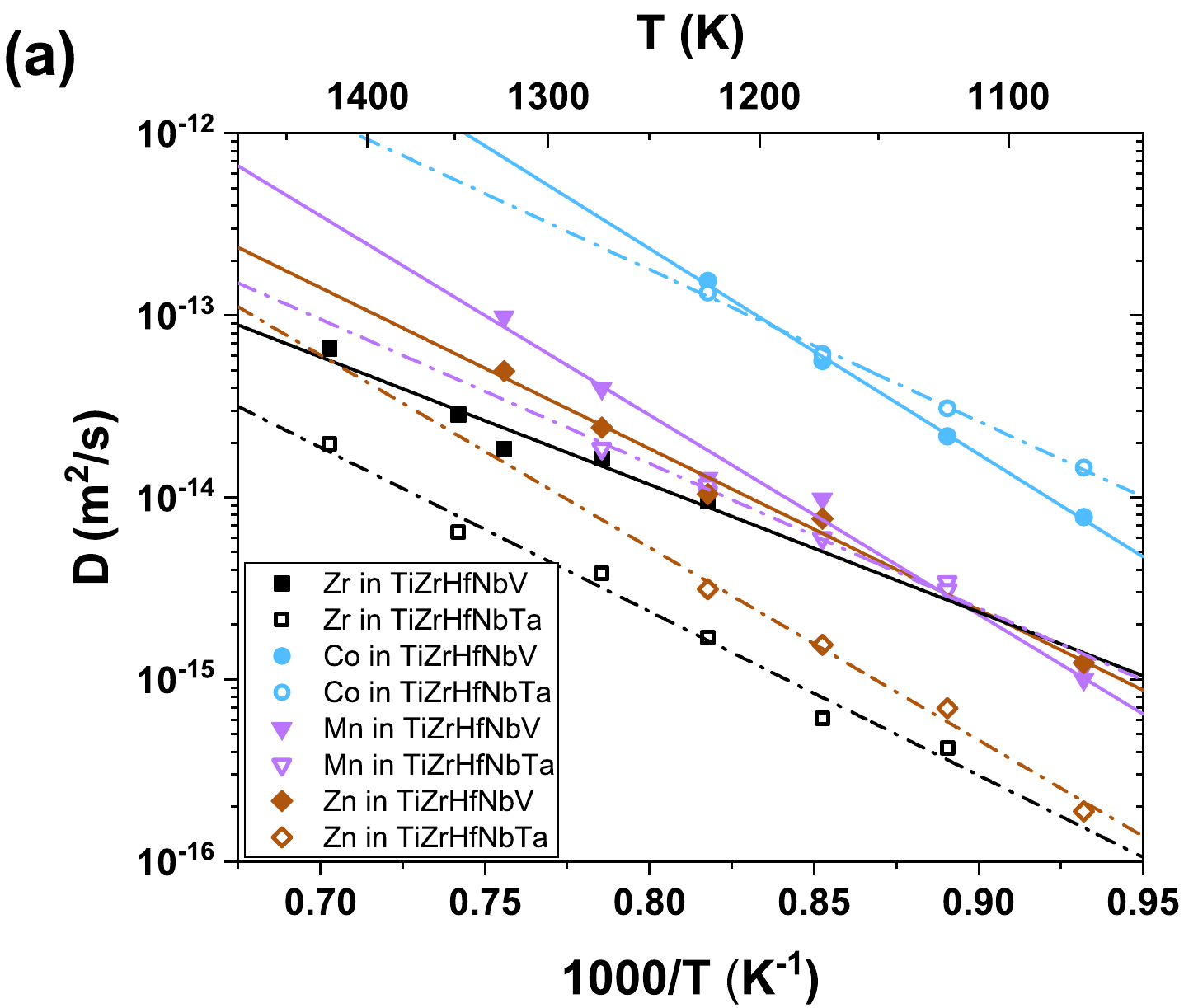}
\hfill
\includegraphics[width=0.495\textwidth]{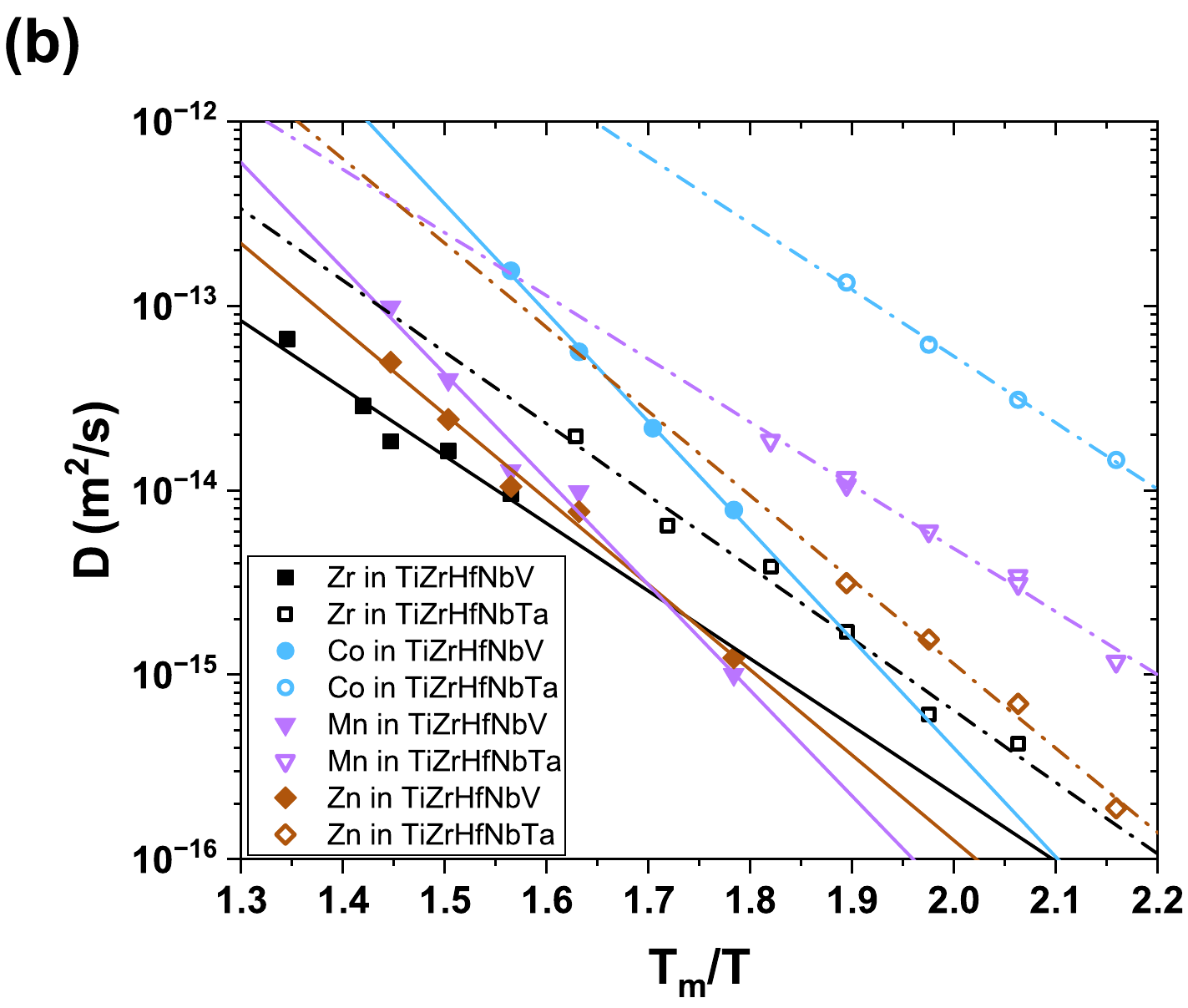}
\caption{Impurity diffusion coefficients of Co (circles), Mn (triangles), and Zn (diamonds), in comparison to self-diffusion coefficients of Zr (squares) \cite{zhang2022zr}, in the TiZrHfNbTa (open symbols) and TiZrHfNbV (filled symbols) RHEAs as a function of (a) the inverse absolute temperature, $T^{-1}$, and (b) the inverse homologous temperature, $T_{\mathrm{m}}/T$.}
\label{fig:CoMnZnZr}
\end{figure*}

The experimentally determined impurity tracer diffusion coefficients of Co ($D_{\rm Co}$), Mn ($D_{\rm Mn}$), and Zn ($D_{\rm Zn}$) in the TiZrHfNbV and TiZrHfNbTa RHEAs are shown in Figs.~\ref{fig:CoMnZnZr}(a) and (b) as a function of the inverse absolute, $T^{-1}$, and the inverse homologous, $T_{\mathrm{m}}/T$, temperatures, respectively. Here $T_{\mathrm{m}}$ is the melting point of the corresponding alloys. Melting points of 2317~K for TiZrHfNbTa and 1914~K for TiZrHfNbV RHEAs were used \cite{zhang2022zr}. For comparison, the self-diffusion coefficients of Zr ($D_{\rm Zr}$) \cite{zhang2022zr} are shown, too. In this paper, we are using the term ``self-diffusion'' for Zr tracer diffusion in homogeneous Zr-containing alloys following Mehrer's nomenclature \cite{Mehrer1990-sd}.

Among the studied elements, Co consistently exhibits the highest diffusion coefficients in both RHEAs across the whole temperature range on the absolute temperature scale, Fig.~\ref{fig:CoMnZnZr}(a). The activation energy, $Q$, of Co diffusion in TiZrHrNbV is the largest ($Q_{\rm Co}$$>$$Q_{\rm Mn}$$>$$Q_{\rm Zn}$$>$$Q_{\rm Zr}$), while in TiZrHfNbTa, it is quite small ($Q_{\rm Zn}$$>$$Q_{\rm Zr}$$>$$Q_{\rm Co}$$>$$Q_{\rm Mn}$). A cross-over temperature ($T_\textrm{cr}$) within the investigated temperature range is observed, above which the diffusion rate of Co in TiZrHfNbV is higher than that in TiZrHfNbTa. The value of $T_\textrm{cr}$ for Co diffusion is around 1190 K, seen from Fig.~\ref{fig:CoMnZnZr}(a). The activation energy of impurity diffusion of Co in TiZrHrNbV (215.4~kJ/mol) is higher than that in TiZrHfNbTa (159.6~kJ/mol), which is different from self-diffusion of Zr (134.2~kJ/mol and 172.4~kJ/mol for TiZrHrNbV and TiZrHrNbTa, respectively). The pre-exponential factor, $D_0$, for Co diffusion in TiZrHfNbV ($2.5^{+2.1}_{-1.1}\cdot 10^{-4}$~m$^2$/s) is higher than in TiZrHfNbTa ($8.3^{+6.9}_{-3.8}\cdot 10^{-7}$~m$^2$/s), too. In comparison to the Arrhenius parameters, particularly pre-exponential factors, of the self-diffusion rates of Zr ($4.77\cdot 10^{-9}$~m$^2$/s in TiZrHfNbV and $3.77\cdot 10^{-8}$~m$^2$/s in TiZrHfNbTa), the $D_0$ values for Co diffusion in these two RHEAs are much larger. 

In the TiZrHfNbV RHEA, impurity diffusivities of Mn and Zn, as well as the self diffusivity of Zr are approximately similar at around 1114~K. Diffusion rates of Zr, Mn, and Zn are one order of magnitude lower than the diffusion rate of Co at $T_\textrm{cr}$, the above-mentioned cross-over temperature for Co. It is worth mentioning that the diffusion rates follow the order: $D_{\rm Co}$$>$$D_{\rm Mn}$$>$$D_{\rm Zn}$$>$$D_{\rm Zr}$ in the range above $T_\textrm{cr}$. The differences in the pre-exponential factors, $D_0$, are important in this case where higher activation energies correlate with faster diffusion. The interplay between the diffusion rate and the activation energy presents an interesting phenomenon for impurity diffusion in the TiZrHfNbV RHEA. The largest activation energy of Co seems to correlate with the most negative mixing enthalpy between Co and the constituent alloying elements, see Table~SI in Supplementary Materials. However, the activation enthalpy of Co diffusion is larger in TiZrHfNbV with respect to that in TiZrHfNbTa, but an opposite trend would be suggested considering the mixing enthalpies. On the other hand, Co might facilitate easier diffusion due to its smallest atomic radius (Co: 1.25 \AA, Zn: 1.34 \AA, Mn: 1.39 \AA, Zr: 1.60 \AA). This indicates the possibility of alternative diffusion mechanisms (to the simple vacancy-mediated one), which will be discussed in Section~\ref{StandardError}.


In the TiZrHfNbTa RHEA, the self-diffusion of Zr is approximately one order of magnitude slower than the impurity diffusion of Mn and two orders of magnitude slower than that of Co on the inverse absolute temperature scale. The impurity diffusion of Zn falls between Zr and Mn within the studied temperature range. A similar trend is observed in the TiZrHfNbV RHEA, where $D_{\rm Co}$$>$$D_{\rm Mn}$$>$$D_{\rm Zn}$$>$$D_{\rm Zr}$. However, the activation energy does not follow this diffusion coefficient trend. The activation energies for Co and Mn are relatively low and close to each other, while that for Zr is higher, with the diffusion of Zn having the largest activation energy.

A comparative analysis of the two RHEAs reveals that Zr and Zn exhibit faster diffusion with lower activation energy in TiZrHfNbV compared to TiZrHfNbTa in the studied temperature range using the inverse absolute temperature scale. The cross-over temperature, $T_\textrm{cr}$, and a higher activation energy in TiZrHfNbV are observed in the case of Co and Mn. A significantly different relation between all the diffusivities in TiZrHfNbV and TiZrHfNbTa can be observed when compared at the inverse homologous temperatures, Fig.~\ref{fig:CoMnZnZr}(b). Diffusion of all the impurity species and Zr is faster in TiZrHfNbTa than in TiZrHfNbV, although TiZrHfNbV has a lower melting point.

\subsection{Diffusion mechanism of Co in the RHEAs and other BCC matrices}\label{StandardError}

Co is technologically relevant due to its widespread application in Co-based hardmetals and alloys with refractory elements. As a result, impurity diffusion of Co has been measured in many BCC refractory metals, especially those, constituting the present HEAs, namely $\beta$-Ti \cite{gibbs1963diffusion}, $\beta$-Zr \cite{kidson1969self}, Nb \cite{pelleg1976diffusion}, V \cite{pelleg1975diffusion}, and Ta \cite{1977diffusion}, as well as BCC binary systems Zr--50\%Ti\cite{herzig1983comment} and Zr--32\%Nb\cite{herzig1987fast}. In $\beta$-Zr, Nb, and Ta, Co shows diffusion rates by two or more orders of magnitude higher than the corresponding self-diffusion rates. The impurity diffusion of Co in $\beta$-Ti is by 1.5 orders of magnitude faster than the self-diffusion rate of Ti, while in V it is by an order of magnitude faster. A similar tendency is also evident in our results for Co diffusion in the TiZrHfNbV and TiZrHfNbTa RHEAs. Mn and Zn show qualitatively similar diffusion behavior, with the diffusion rates being somewhat enhanced with respect to those measured for Zr self-diffusion, Fig.~\ref{fig:CoMnZnZr}. It was speculated that Co diffusion in the above-mentioned BCC metals/alloys is not governed purely by the vacancy mechanism \cite{kidson1969self, pelleg1976diffusion, wenwer1989two}, which is typical for substitutional diffusion in metals \cite{Mehrer2007}. For example, a thermodynamic model of vacancy-mediated diffusion in BCC metals elaborated by Neumann et al. \cite{neumann1997modified} revealed a perfect agreement for the majority of investigated impurities in W, while the predicted diffusion rates of Co in W were significantly underestimated with respect to the experimental values. This fact was interpreted \cite{wenwer1989two, neumann1997modified} as an indication for a dissociative diffusion mechanism \cite{frank1956mechanism} for Co in W. In this case, the impurity atoms, occupying predominantly the substitutional lattice sites, jump occasionally to a neighboring interstitial position leaving a vacancy behind and diffuse fast recombining eventually with a further vacancy (or returning back) \cite{Stolwijk1990}. 

The presence of impurities is of particular importance since their strong interactions with structural defects modify vacancy-atom interactions and the corresponding jump barriers. Following the original Howard and Lidiard's 5-jump frequency model for the FCC lattice\cite{howard1964-5jump}, multi-frequency models were developed to account for different jump frequencies of the vacancies due to the presence of impurity atoms in different crystalline lattices\cite{le1978solute,lidiard1997atomic}. Within the four-frequency model for a BCC lattice, the diffusion coefficients of the matrix (solvent) and impurity atoms are
\begin{equation}
D_\text{solvent} = a^2 \omega_0 f_0 C^\text{eq}_\text{VA}
\end{equation}
and
\begin{equation}
D_\text{impurity} = a^2 \omega_2 f_2 \frac{\omega_4}{\omega_3} C^\text{eq}_\text{VA},
\end{equation}
respectively. Here, \(a\) is the lattice parameter, \(f_0\) is the self-diffusion correlation factor (0.727 for the BCC lattice), and \(f_2\) is the impurity correlation factor for dilute BCC alloys\cite{le1978solute}. Further, \(C^\text{eq}_\text{VA}\) is the equilibrium vacancy concentration, and \(\omega_i\) is the atom-vacancy exchange frequency. The relevant frequencies are defined as:
\begin{itemize}
     \item \(\omega_0 \): The jump frequency of host lattice atoms when the vacancy is far from the impurity atom.
     \item \(\omega_2\): The frequency of vacancy--impurity exchanges in the nearest-neighbor configurations.
    \item \(\omega_3\): The frequency of a vacancy escaping from a nearest-neighbor position around the impurity atom (dissociative jump for the vacancy--impurity pair).
    \item \(\omega_4\): The frequency of vacancy jumps opposite to $\omega_3$ (associative jumps for the vacancy--impurity pair).
\end{itemize}

The solvent jump frequency, $\omega_0$, enters into $D_{\rm impurity}$ through the correlation factor $f_2$. In simple cases, like vacancy diffusion in isotropic crystals, $f_2$ has the form \cite{le1978solute}
\begin{equation}\label{eq:f2}
f_2 = \frac{H}{(2\omega_2 + H)},
\end{equation}
where $H$ is a function of \emph{all} vacancy jump frequencies. 

We may adopt this model for the present case of impurity diffusion in RHEAs, treating the vacancy jump frequencies as \textit{effective} values that are averaged to account for the diverse chemical environments in multi-principal element alloys. Within this approximation, the ratio of the Co (impurity) and Zr (considered as an effective host) diffusion coefficients reads
\begin{equation}
\label{eq:ratio}
    \frac{D_\textrm {Co}}{D_\textrm{Zr}} = \frac{\omega_2}{\omega_0} \frac{\omega_4}{\omega_3}\frac{1.376H}{2\omega_2+H},
   \end{equation}
\begin{equation}
    H = {\omega_3}F\left(\frac{\omega_4}{\omega_0}\right),
\end{equation}
where $F$ varies between 3 and 7 depending on the $\omega_4/\omega_0$ ratio \cite{le1978solute}.

In the case of the TiZrHfNbTa RHEA, Co diffusion is significantly faster than the self-diffusion rate of Zr by two or even more orders of magnitude. Such an enhancement of impurity diffusion is not compatible with the vacancy-mediated mechanism. For example, if $\omega_2$ exceeds other frequencies by a factor larger than 100, the diffusion enhancement, $D_\textrm {Co}/D_\textrm{Zr}$, will be proportional to $\omega_4/2\omega_0$ (with a pre-factor of about 10) and $\omega_2$ in Eq.~\eqref{eq:ratio} will effectively be canceled. The impurity performs highly correlated jumps, corresponding to the small values of the correlation factor $f_2$, see Eq.~\eqref{eq:f2}.  Thus, alternative diffusion mechanisms have to be analyzed.

\begin{figure*}
\centering 
\includegraphics[width=0.9\linewidth]{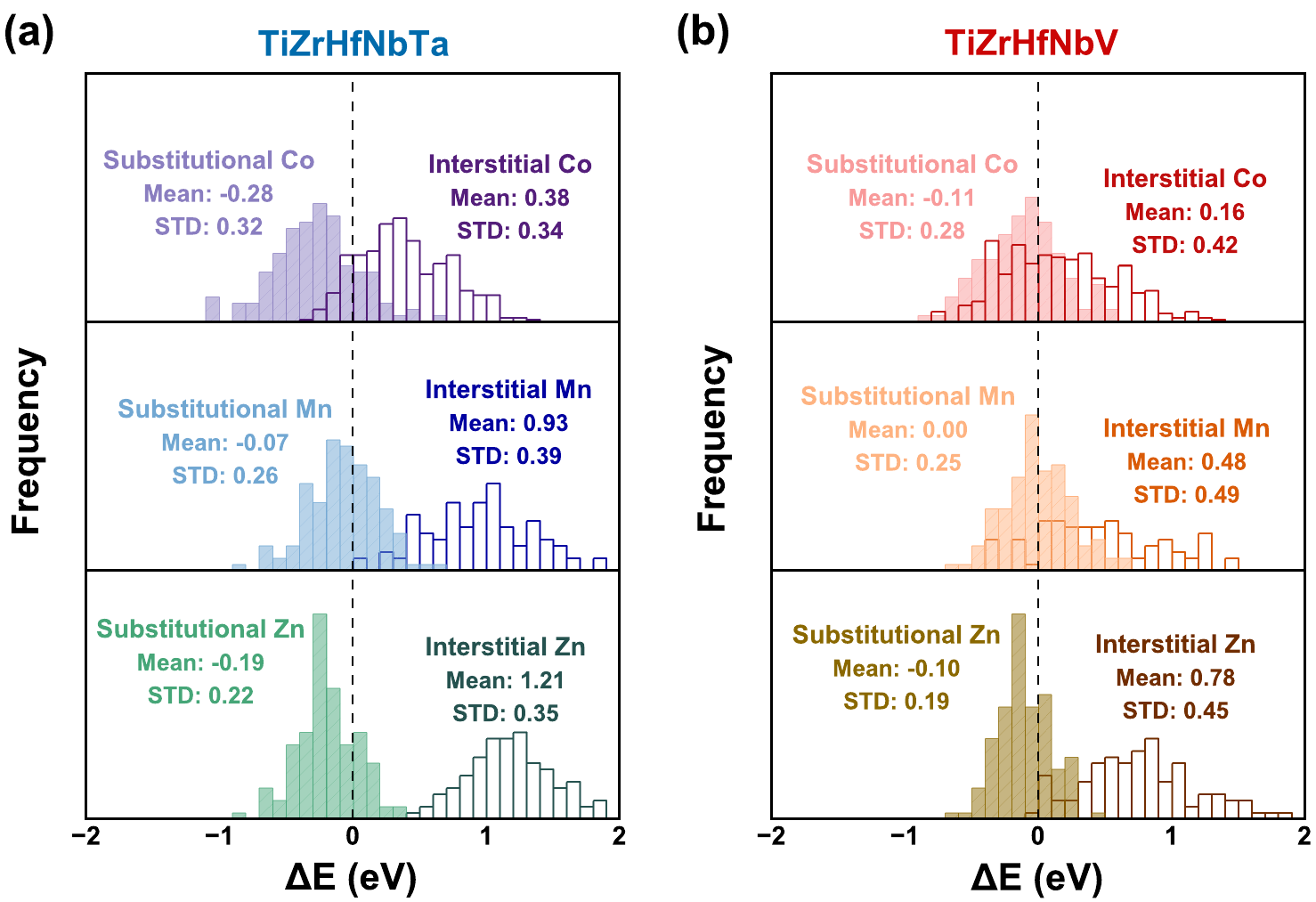}
\hfill
\caption{Distributions of substitutional and interstitial solution energies of Co, Mn, and Zn in the (a) TiZrHfNbTa and (b) TiZrHfNbV RHEAs. The mean value and the corresponding standard deviation (in eV) are given for each distribution.}
\label{fig:DFT_sol}
\end{figure*}

In order to provide further insights into the diffusion mechanisms of solutes in RHEAs, we performed an \textit{ab initio}-based analysis of Co, Mn, and Zn solution energies at both substitutional as well as interstitial positions of the BCC lattice.

In  Figs.~\ref{fig:DFT_sol}(a) and (b), the spectra of the solution energies for substitutional and interstitial positions for Co, Mn, and Zn atoms in TiZrHfNbTa and TiZrHfNbV are shown as histograms. 
A more negative (or more positive) solution energy at elevated temperatures corresponds to a higher (or lower) probability of the impurity occupying a given site. Additionally, diffusion through interstitial sites is generally considered to be much faster than through substitutional sites. This is because diffusion via substitutional sites relies heavily on the availability of vacancies, which are typically present in very small concentrations, and typically larger migration barriers are involved in the atom--vacancy exchanges with respect the migration barriers between interstitial sites. Low substitutional solution energies dominate for these solutes in both RHEAs. However, there are notable qualitative differences between substitutional and interstitial solution energies for both RHEAs and impurities, particularly concerning Co. The average interstitial solution energy of Co is $0.38 \pm 0.34$ eV in TiZrHfNbTa and $0.16 \pm 0.42$ eV in TiZrHfNbV. The standard deviations indicate a significant proportion of attractive interstitial sites with large negative solution energies, especially notable in TiZrHfNbV, as illustrated in Figs. \ref{fig:DFT_sol}(b), top panel. This finding aligns with the experimental data showing rapid diffusion of Co.

In contrast, Zn exhibits a qualitatively different behavior. In TiZrHfNbTa, the interstitial solution energies are larger than the substitutional ones. In TiZrHfNbV though, a considerable fraction of interstitial solution energies are comparable to substitutional energies. This observation is consistent with the experimental results, as Zn diffuses much faster in TiZrHfNbV than in TiZrHfNbTa, as shown in Fig. \ref{fig:CoMnZnZr}(a).

Mn shows a mixed behavior. In TiZrHfNbTa, Mn behaves similarly to Zn, although the interstitial and substitutional solution energies slightly overlap. In TiZrHfNbV, the interstitial solution energies are significantly lower. However, the reduction is less pronounced compared to Co. Overall, the trends in substitutional and interstitial solution energies correlate with the measured diffusion rates for Co, Mn, and Zn in both RHEAs.

These conclusions can be drawn, even though demanding calculations of the energy barriers and attempt frequencies for a huge number of different chemical configurations with subsequent kinetic Monte Carlo simulations are required for an ultimate comparison. The DFT-informed calculations seem to rule out the interstitial diffusion mechanism for the Zn atoms in the TiZrHfNbTa RHEA and, at least, allow its contribution in TiZrHfNbV. For the Co atoms, the situation is different. A strong contribution of the interstitial mechanism is elaborated for Co in RHEAs, in good agreement with the present experimental data.

\begin{figure}[ht]
\centering 
\includegraphics[width=0.9\linewidth]{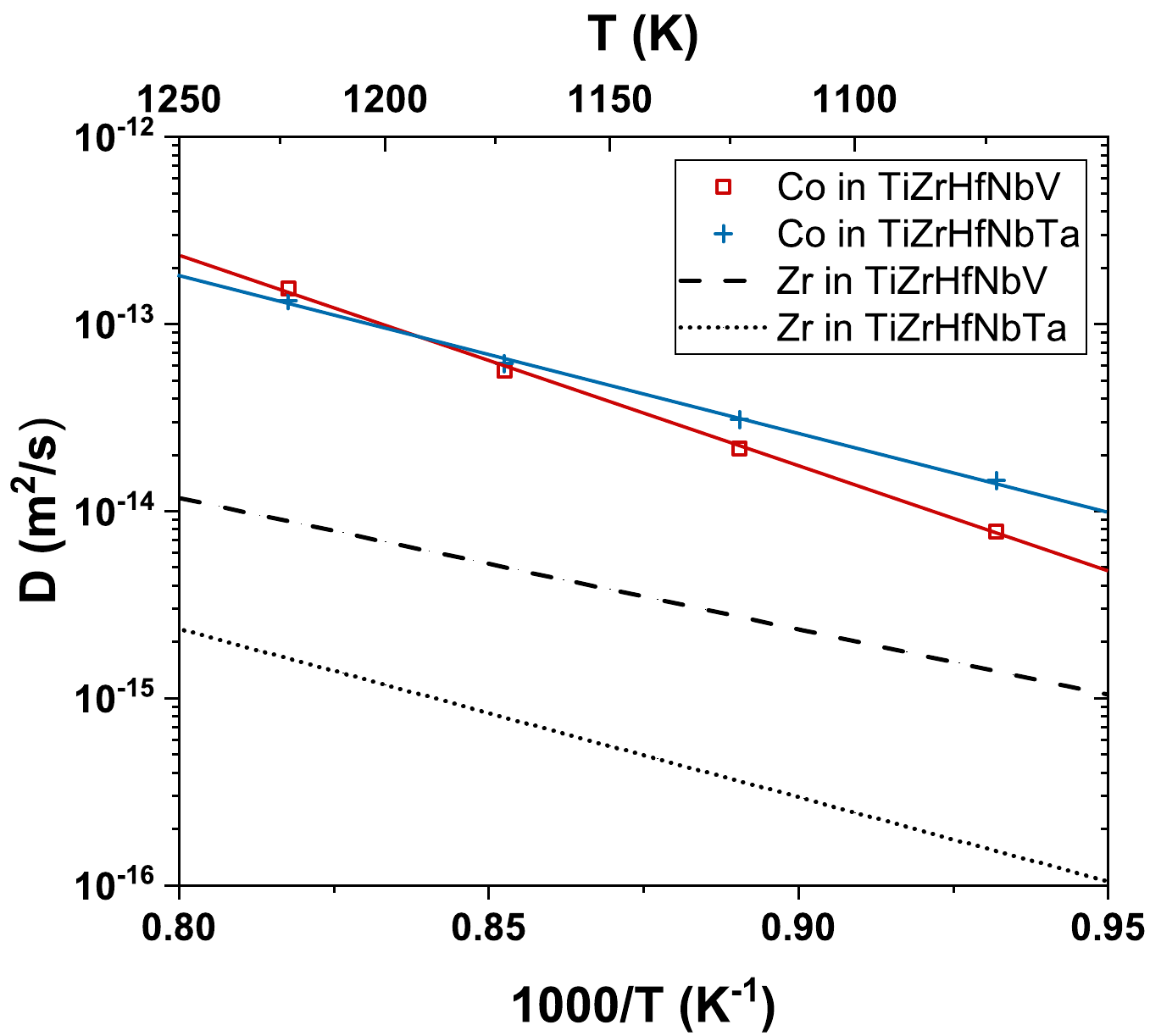}
\caption{Impurity diffusion coefficients of Co, $D_{\rm Co}$, compared to self-diffusion coefficients of Zr, $D_{\rm Zr}$, in the two investigated RHEAs.}
\label{fig:CoZr}
\end{figure} 

Indeed, Co diffusion in TiZrHfNbTa is significantly faster than the self-diffusion rate of Zr by two or even more orders of magnitude as shown in Fig. \ref{fig:CoZr}. The enhancement of Co diffusion in TiZrHfNbV is smaller, generally within the bounds imposed by the vacancy-mediated diffusion mechanism in the BCC lattice \cite{Mehrer2007, PETERSON1969}. Nevertheless, guided by the DFT-informed analysis, we suggest that the dissociative mechanism is responsible for the diffusion of Co in TiZrHfNbV RHEA, too. 

The calculated solution energies indicate the partition of the impurity atoms between two types of positions, though the degree of the enhancement is also affected by the migration barriers. In particular, the diffusion coefficient of an impurity $Y$ according to the dissociative diffusion mechanism is, see e.g. \cite{Herzig2002},
\begin{equation}
    D_Y=\frac{C^{\rm int}_Y}{C^{\rm int}_Y+C^{\rm sub}_Y}D^{\rm int}_Y+\frac{C^{\rm sub}_Y}{C^{\rm int}_Y+C^{\rm sub}_Y}D^{\rm sub}_Y.
\end{equation}
Here, $C^{\rm int}_Y$ and $C^{\rm sub}_Y$ are the concentrations of the $Y$ atoms in interstitial and substitutional positions, and $D^{\rm int}_Y$ and $D^{\rm sub}_Y$ are the respective diffusion coefficients. Thus, large interstitial solubility and presumably low migration barriers for the interstitial jumps contribute to the enhanced diffusivity of Co in RHEAs.

\begin{figure*}
\centering 
\includegraphics[width=0.49\textwidth]{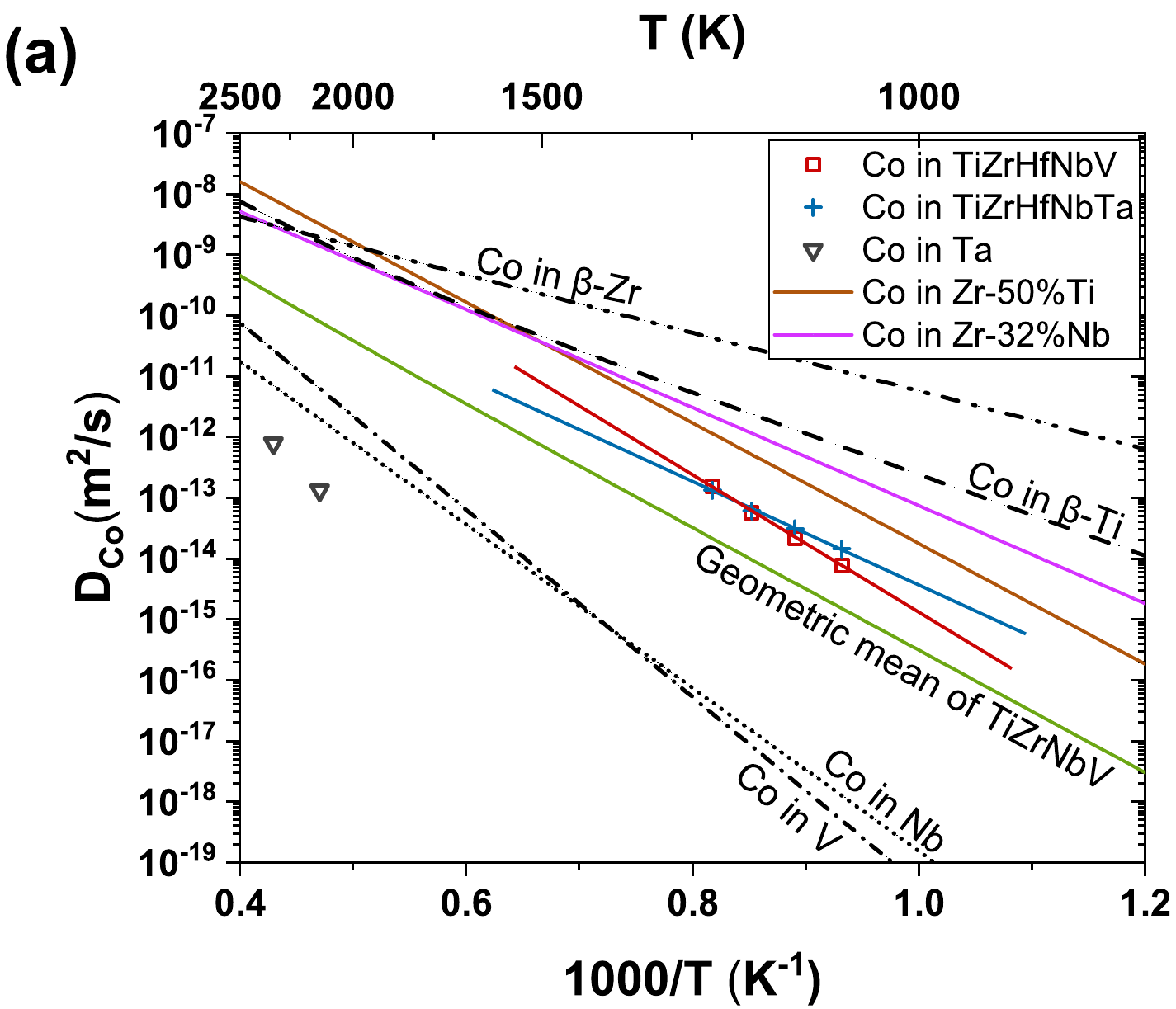}
\hfill
\includegraphics[width=0.49\textwidth]{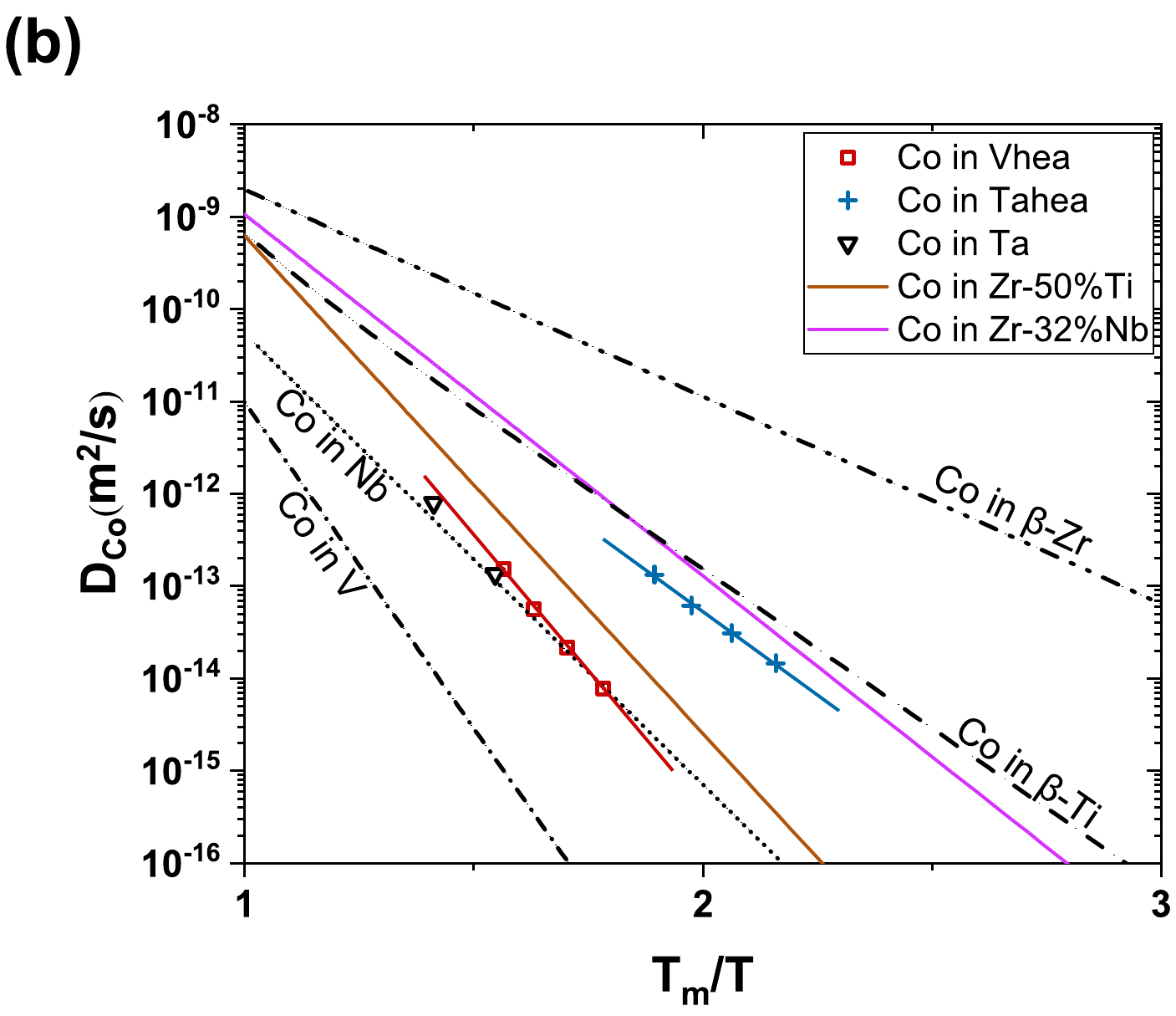}

\caption{Co diffusion coefficients, $D_{\rm Co}$, in the two investigated RHEAs compared to Co diffusion coefficients in the BCC phases of the respective pure metals ($\beta$-Ti \cite{gibbs1963diffusion}, $\beta$-Zr \cite{kidson1969self}, Nb \cite{pelleg1976diffusion}, V \cite{pelleg1975diffusion}, and Ta \cite{1977diffusion}), as well as binary subsystems Zr--50\%Ti\cite{herzig1983comment} and Zr--32\%Nb\cite{herzig1987fast} as a function of (a) the inverse absolute, $T^{-1}$, and (b) the inverse homologous, $T_{\mathrm{m}}/T$, temperature. In (a), the geometric mean of the diffusion coefficients in pure elements is shown for comparison.}
\label{fig:Co}
\end{figure*} 

\subsection{Tracer diffusion in RHEAs: chemical complexity vs. lattice distortions}
In Fig.~\ref{fig:Co}(a), the Co tracer diffusion rates in BCC pure elements Ti, Zr, Nb, V, and Ta are compared using the inverse absolute temperature scale. In the pure BCC metals, the Co diffusivities within the temperature range of the present 
measurements follow the order $D_{\rm Zr}$ $>$ $D_{\rm Ti} > D_{\rm Nb} > D_{\rm V} > D_{\rm Ta}$, which is not consistent with the relation between the melting points of these elements. Apparently, diffusion in group V (V, Nb, and Ta) elements is much slower with a much larger activation energy than in group IV (Ti and Zr) elements. Co diffusion in two binary alloys is also shown in Fig.~\ref{fig:Co}(a). For the Zr--50\%Ti alloy, a pattern in the variability of the diffusion rate of Co is difficult to identify because of the relatively slow diffusion in the binary as compared to pure Ti and Zr. For the Zr--32\%Nb alloy, the diffusion rate of Co is more predictable, falling between that of pure Ti and Zr, although it remains slightly faster than in the Zr--50\%Ti alloy. The trend does not change too much when using the inverse homologous temperature scale for comparison (Fig.~\ref{fig:Co}(b)).  

The geometric mean of the self-diffusion rates of the constituting elements of RHEAs was argued to be a representative value for scrutinizing the "sluggish" diffusion concept \cite{daw2021sluggish,sen2023anti, Starikov2024}. Here, we extend the original approach of Daw and Chandross \cite{daw2021sluggish} towards quantifying the impact of the chemical complexity on impurity diffusion with respect to the corresponding diffusivities in unaries. In Fig.~\ref{fig:Co}(a), we added the geometric mean value of Co impurity diffusion in $\beta$-Ti, $\beta$-Zr, Nb, and V (Co diffusion data in $\beta$-Hf is unavailable), i.e., the constituting elements of TiZrHfNbV. The existing diffusion data for Ta are too limited to perform a similar analysis. However, it is reasonable to assume that the geometric mean from the unaries for TiZrHfNbTa will fall below the geometric mean for TiZrHfNbV. Thus, the comparison reveals a significant enhancement of the Co diffusion rate with respect to the ``averaged'' diffusion rate in the RHEAs under investigation, by an order of magnitude within the experimental temperature range. Moreover, we may reasonably assume that Co diffusion in $\beta$-Hf is quite slow because of the high melting point of Hf. 

Daw et al.\cite{daw2021sluggish} concluded that the lattice mismatch, $\delta$ (as defined in Eq.~(\ref{eq:delta})), plays a dominant role in governing the self-diffusion behavior in multi-principal element alloys. Specifically, sluggish self-diffusion is more likely to occur in alloys with a small lattice mismatch (1–3\%), whereas alloys with a larger lattice mismatch ($\delta > 3$\%) tend to exhibit significantly faster self-diffusion. This trend is consistent with the findings of the present study for impurity diffusion, where Co exhibits faster diffusion in the two RHEAs compared to the corresponding geometric mean. However, the larger local lattice distortions in the TiZrHfNbV RHEA do not lead to more enhanced Co diffusion compared to the TiZrHfNbTa RHEA, indicating probably the influence of another, not purely vacancy-mediated mechanism. We argue that this is the dissociative diffusion mechanism, which governs fast diffusion of Co in RHEAs. Other factors such as thermal anharmonic vibrations may also play a critical role in the enhancement of diffusion, as revealed by a recent \textit{ab initio} study for BCC W (also a 5$d$ element in the same row as Ta in the periodic table)~\cite{Zhang2025}.

The local lattice distortions (as provided by the MSDs, see Fig.~\ref{fig:MSD}) are larger in TiZrHfNbV than in TiZrHfNbTa. When viewed on the inverse absolute temperature scale, it is hard to draw any definite correlation between local lattice distortions and the impurity diffusion rates due to the mentioned cross-over effect for Co and Mn diffusion. On the inverse homologous temperature scale, a certain deceleration in Co diffusion (as impurity) is evident in the more locally distorted TiZrHfNbV RHEA. Similar trends were observed for self-diffusion in the FCC HEAs of the CoCrFeMnNi family, where self-diffusivities decrease compared to the less complex CoCrFeNi HEA\cite{vaidya2018bulk}, or conventional dilute solid solutions such as Co$_{10}$Cr$_{10}$Fe$_{10}$Mn$_{10}$Ni$_{60}$ and Co$_{2}$Cr$_{2}$Fe$_{2}$Mn$_{2}$Ni$_{92}$\cite{kottke2019tracer,kottke2020experimental} when analyzed using the inverse homologous temperature scale. In these cases, the local lattice distortions increase with increasing mixing entropy, which may contribute to the decelerated diffusion on the inverse homologous temperature scale. In the case of impurity diffusion in the BCC HEAs, opposite effects can probably be expected. For example, diffusion retardation is seen for Co (interstitial solubility) and an acceleration of diffusion
for Mn correlates with its substitutional solubility.

Wang et al.\cite{wang2020effect} evaluated indirectly the diffusion rates of oxygen in the TiZrNbTa RHEA, where two distinct oxidation phenomena were observed. At lower temperatures, more severe lattice distortions in the TiZr-rich region resulted in faster oxygen diffusion and higher oxidation rates, leading to significant stress, crack formation in the oxide layer, and ultimately catastrophic pesting\cite{wang2020effect}. At higher temperatures, oxygen diffusion becomes more homogeneous, resulting in parabolic oxidation behavior and a thicker oxide layer. 


Accounting for the present results and the literature data, we conclude that the severe lattice distortions and the chemical complexity of RHEAs give rise to multifold effects and cannot be simply described by a single concept, aka "sluggish" diffusion.

\subsection{Impact of volume expansion and short-range order}\label{sec:volume_SRO}

\begin{figure*}[ht]
\centering 
\includegraphics[width=1.0\linewidth]{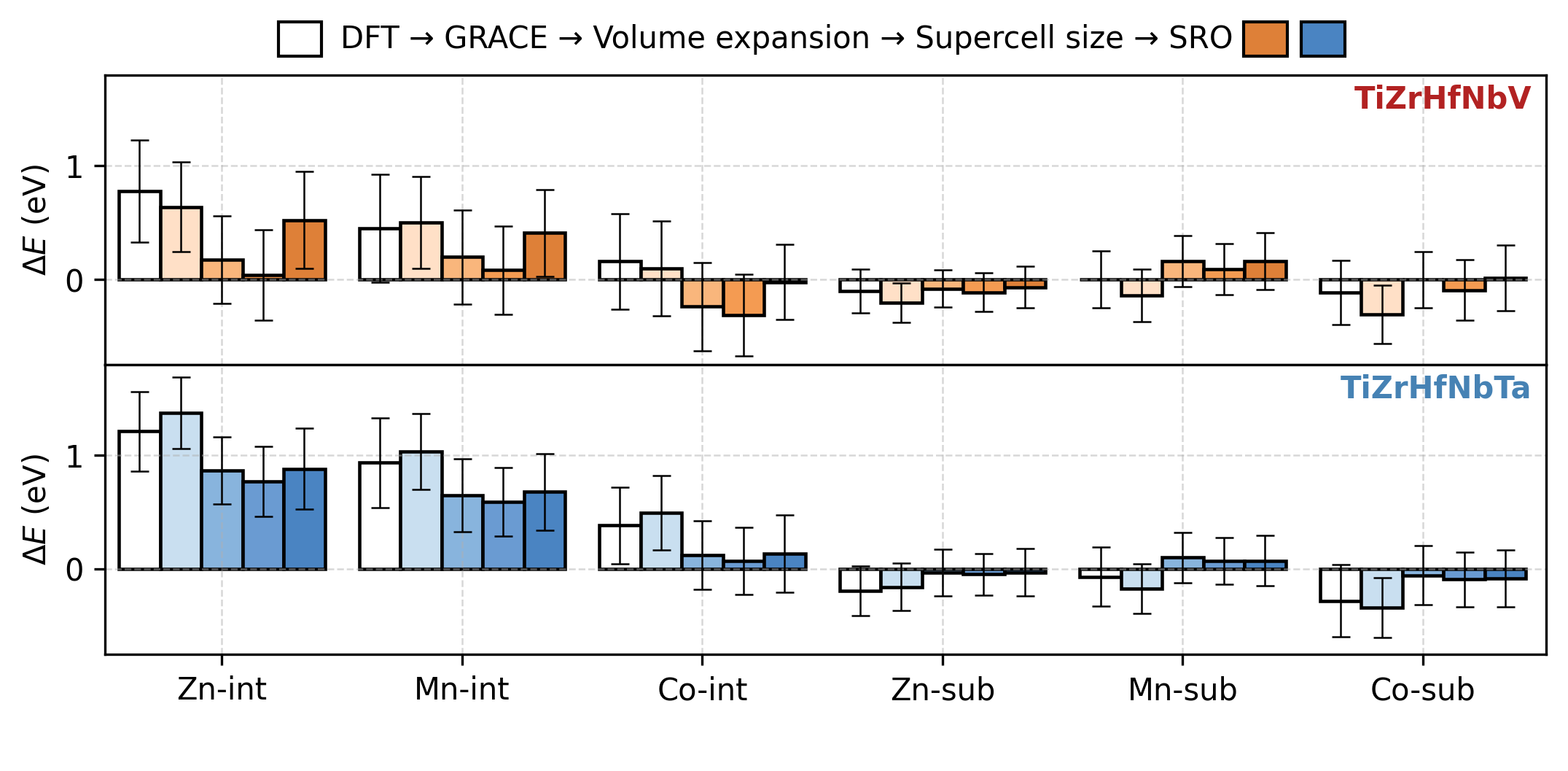}
\caption{Mean solution energies and standard deviations for Zn, Mn, and Co solutes in the TiZrHfNbV (top) and TiZrHfNbTa (bottom) alloys. Five configurations are compared for each solute: direct DFT (white), GRACE with universal potentials (light color), inclusion of thermal volume expansion, larger supercell size, and short-range order (last column,  darkest color). The progression highlights the impact of volume, finite size, and SRO effects on defect energetics.}
\label{fig:GRACE}
\end{figure*}

The foregoing diffusion analysis has been done assuming a completely random solution model. This approximation is thought to be well suitable for the present case of diffusion in RHEAs at relatively high temperatures. 
According to CALPHAD calculations \cite{LI2023-calphad}, below about 1100~K both RHEAs decompose to mixtures of two BCC-based phases. Within the temperature interval of the present study, i.e. $1073$~K~$\le T \le1323$~K, no phase separation was observed, at least for the relatively short annealing times of several days. Theoretically, one might expect development of SRO at lower temperatures and the presence of the short-range order might influence the measured diffusion rates. Additionally, the DFT analysis was conducted without accounting for temperature-induced volume expansion of the alloys. Besides SRO, this temperature effect could further alter the computed solution energies.

To evaluate the influence of volume expansion at finite temperatures, finite supercell size, and SRO on solute energetics, we utilized machine-learning interatomic potentials derived from foundational models, commonly referred to as universal potentials (see Sec.~\ref{Sec:Method_SRO}). This approach allowed for a comprehensive assessment of the different physical effects without resorting to further computationally intensive DFT calculations.

To facilitate direct comparisons across the different scenarios, the computed mean solution energies and their associated standard deviations are shown in Fig.~\ref{fig:GRACE}. The standard deviation serves as a qualitative indicator of the spread in the solution energies.

A direct comparison between the explicit DFT results and those obtained from the universal machine-learning potentials is also shown in Fig.~\ref{fig:GRACE}. The first two bars for each solute-defect combination represent the DFT values (white bars) and the corresponding GRACE predictions for the same settings (the second bars). A good qualitative agreement is observed across all solutes and defect types, demonstrating that the universal potentials reliably reproduce the DFT trends.

We first assessed the influence of thermal volume expansion. While the DFT calculations were carried out at the equilibrium lattice constants (0 K), we used a quasi-harmonic Debye model to estimate the lattice parameters at 1000 K and recalculated the defect energetics using GRACE. The resulting values, represented by the third bars, indicate a significant reduction in interstitial solution energies. This suggests that inserting an interstitial solute at 0 K creates internal pressure, which is alleviated at the expanded high-temperature volume. In contrast, the impact on substitutional defects is smaller and of the opposite sign, indicating a weakening of chemical bonding at the larger volume.

To explore potential supercell size effects, we compared results from 128-atom and 250-atom supercells, which are represented as the third and fourth bars, respectively. Only marginal differences were observed, implying that the smaller supercells already provide sufficient accuracy to capture the relevant trends for the defect energetics.

As outlined in Sec.~\ref{Sec:Method_SRO}, we further investigated the role of SRO by performing Monte Carlo simulations using on-lattice low-rank potentials fitted to the GRACE data. We extracted representative supercells from these simulations, including SRO at 1000~K, and recalculated the defect solution energies. These results are shown as the final bars in each group in Fig.~\ref{fig:GRACE}. Compared to the random configurations, SRO results in only minor changes to the defect energies, with the most pronounced effect observed for interstitials in the TiZrHfNbV alloy, exhibiting slightly stronger SRO than TiZrHfNbTa.

Overall, by comparing the initial DFT results (first white bars) with the GRACE results incorporating finite temperature volume expansion, supercell size, and SRO (final dark bars), we conclude that the trends in relative defect energetics across solutes remain qualitatively robust. While including thermal and chemical disorder effects refines the absolute values, it does not alter the fundamental ordering or behavior of the considered solutes in the two alloys.

In addition to defects' energetics, SRO can influence the vacancy concentration and diffusion barriers in a multi-principal element alloy. 
Recent theoretical studies have shown that the extent to which SRO influences diffusion behavior varies with system, temperature, and specific alloy chemistry. SRO induces preferential vacancy diffusion pathways and might give rise to non-monotonic variations in the diffusion coefficients, e.g. in CoCrFeNi HEAs at 500--700 K \cite{IBRAHIM2024155335}. It also reduces diffusivity by increasing migration energy barriers, enhancing diffusion correlation, and promoting vacancy localization \cite{XING2022114450}. Bidirectional coupling exists between self-diffusion and SRO evolution: self-diffusion drives SRO formation, while SRO further suppresses long-range diffusion \cite{PhysRevMaterials.7.033605}.

Within the accuracy of the present tracer diffusion measurements, no deviations from the Arrhenius-type temperature dependencies were observed, Fig.~\ref{fig:CoMnZnZr}. Thus, the diffusion measurements support the DFT-based analysis that within the high-temperature interval (1073--1323 K), SRO formation is not pronounced in the present RHEAs and it does not influence the solute diffusion behavior in a measurable manner. Therefore, under such conditions, the SQS approach, used in the present paper, is a valid method for capturing the averaged diffusion behavior.
\color{black}

\section{Conclusions}
For the first time, impurity diffusion of Co, Mn, and Zn was measured in the TiZrHfNbV and TiZrHfNbTa BCC RHEAs utilizing suitable radioactive isotopes. Moreover, lattice distortions in these RHEAs were quantitatively assessed using neutron total scattering and first-principles calculations. The main results are summarized as follows:
\begin{itemize} 

\item Both RHEAs exhibit significant local lattice distortions, stronger than in conventional refractory alloys. Compared to TiZrHfNbTa, the TiZrHfNbV RHEA exhibits both larger overall lattice distortions calculated through atomic size mismatch and local lattice distortions characterized by pair distribution functions and mean squared displacements. The larger distortions stem from the relatively small atomic radius of V.

\item Co consistently exhibits the highest diffusion coefficients across both RHEAs for the entire temperature range. The diffusion rates of Mn, Zn, and Zr are significantly lower than that of Co in both alloys, with $D_{\rm Co}$$>$$D_{\rm Mn}$$>$$D_{\rm Zn}$$>$$D_{\rm Zr}$ for the entire studied temperature range in the TiZrHfNbTa RHEA and for temperatures higher than a cross-over temperature (about 1125K) in TiZrHfNbV.

\item Interstitial jumps of Co atoms in terms of the dissociative mechanism are expected to govern Co diffusion in both alloys, especially in the TiZrHfNbTa RHEA, as suggested by the DFT-based calculations of Co solubility.

\item Lattice distortions are found to contribute to the measured enhancement of Co diffusivity with respect to the anticipated 'averaged' behavior which may be predicted by analyzing the diffusion rates of Co in pure elements.

\item A potential impact of SRO on solubility of impurity atoms in both RHEAs has been found to be small and unlikely to influence the observed diffusion behavior in the temperature interval under investigation.
\end{itemize}

\section*{CRediT authorship contribution statement}
Jingfeng Zhang: Writing -- original draft, Visualization, Investigation, Formal analysis, Data curation. Xiang Xu:  Writing -- original draft, Methodology, Data curation, Visualization, Formal analysis. Fritz K$\rm{\ddot{o}}$rmann: Writing -- review \& editing, Methodology, Data curation, Visualization, Formal analysis. Wen Yin: Data curation. Xi Zhang:  Writing -- review, Supervision, Funding acquisition. Christian Gadelmeier: Resources. Uwe Glatzel: Writing -- review \& editing, Supervision. Blazej Grabowski: Writing -- review \& editing, Supervision. Runxia Li: Writing -- review \& editing, Supervision, Funding acquisition. Gang Liu: Writing -- review \& editing, Supervision. Biao Wang: Writing -- review \& editing, Supervision. Gerhard Wilde: Writing -- review \& editing, Supervision. Sergiy V. Divinski: Writing -- review \& editing, Supervision, Funding acquisition, Conceptualization.

\section*{Acknowledgement}
Financial support from the German Research Foundation (DFG) via research grants DI 1419/24-1 (project number 509804947), ZH 1218/1-1 (project number 509804947), GL 181/56-2 (project number 388739063), KO 5080/4-1 (project number 541649719), and  SFB1333 (project ID 358283783-CRC 1333/2 2022), from the Guangdong Basic and Applied Basic Research Foundation (grant number 2023A1515110615), and from the European Research Council (ERC) under the European Union’s Horizon 2020 research and innovation program (Grant Agreement No.\
865855) is acknowledged. We thank the staff members of the Multi-Physics Instrument (https://cstr.cn/31113.02.CSNS.MPI) at the China Spallation Neutron Source (CSNS) (https://cstr.cn/31113.02.CSNS) for providing technical support and assistance in data collection and analysis. We also wish to thank Dr.\ Pengfei Zhou and Dr.\ Shulin Liu (Spallation Neutron Source Science Center, China) for their help with the measurements.
The DFT calculations were performed on the national supercomputer Hawk at the High Performance Computing Center Stuttgart (HLRS) under the grant number H-Embrittlement/44239.
 F.K. acknowledges the LRP and MC simulation packages by Alexander Shapeev. 

%

\bibliographystyle{elsarticle-num}

\bibliography{ref}

\end{document}